\documentclass{article}
\usepackage{graphicx} 
\usepackage{authblk} 
\usepackage{geometry}
\usepackage{placeins}

\usepackage{url}
\usepackage{amsmath}
\usepackage{multirow}
\usepackage{amsmath, amssymb}
\usepackage[dvipsnames,svgnames]{xcolor}
\usepackage[colorlinks=true,
      linkcolor=red,
      urlcolor=blue,
      citecolor=blue]{hyperref}
\usepackage{orcidlink}
\usepackage{float}
\usepackage{tabularx}  
\usepackage{booktabs}  
\usepackage{cite}
\usepackage[utf8]{inputenc}
\DeclareUnicodeCharacter{2212}{-}

\title{Viscous Fluid Models of Cosmic Acceleration in FRW Spacetime Using MCMC Constraints}

\author[1]{Mohit Thakre\orcidlink{0009-0005-1873-8286}}
\author[2]{Praveen Kumar Dhankar\orcidlink{0000-0002-8201-6019}}
\author[3,5,6]{Behnam Pourhassan\orcidlink{0000-0003-1338-7083}}
\author[4]{Safiqul Islam\orcidlink{0000-0003-1373-4137}}

\affil[1,2]{{\scriptsize Symbiosis Institute of Technology, Nagpur Campus, Symbiosis International (Deemed University), Pune 440008, Maharastra, India. $^1$Email: mohitthakre100@gmail.com; $^2$Email: pkumar6743@gmail.com }}

\affil[3]{{\scriptsize School of Physics, Damghan University, Damghan, 3671641167,  Iran. $^3$Email: b.pourhassan@du.ac.ir}}

\affil[5]{{\scriptsize Center for Theoretical Physics, Khazar University, 41 Mehseti Street, Baku, AZ1096, Azerbaijan.}}

\affil[6]{{\scriptsize Centre of Research Impact and Outcome, Chitkara University, Punjab, Rajpura, 140417, India.}}

\affil[4]{{{\scriptsize Department of Basic Sciences, General Administration of Preparatory Year, King Faisal University, P.O. Box 400, Al Ahsa 31982, Saudi Arabia \& Department of Mathematics and Statistics, College of Science, King Faisal University, P.O. Box 400, Al Ahsa 31982, Saudi Arabia. $^4$Email: sislam@kfu.edu.sa}}}

\date{}

\begin{document}
\maketitle

\section*{Abstact}

This study combines theoretical advancements with observational limitations to investigate the cosmological implications of a bulk viscous modified Chaplygin gas (MCG) in a Friedmann--Robertson--Walker (FRW) in (3+1) dimensional spacetime framework.  We provide analytical solutions for both viscous and non-viscous cases, pointing out variations in the energy density evolution, the Hubble parameter dynamics, and the deceleration parameter transitions.  Bulk viscosity suppresses oscillations in structure creation, a well-known drawback of Chaplygin gas models in larger dimensions, as shown by a thorough perturbation analysis. Using the bulk viscosity coefficient and Hubble expansion parameter, which are incorporated by the total pressure and the appropriate pressure and by using energy momentum conservation law determined time-dependent density. With the help of three conditions($\xi = 0$, $\xi\neq0$, and we neglect both bulk viscosity and presence of Chaplygin gas, i.e $A=0$ and $\xi=0$) created three different models as the Hubble parameter is a function of redshift $z$. By applying the MCMC method to these models, we have gone through observational analysis by using the Hubble and Pantheon Type Ia Supernova dataset.

\section{Introduction}
Characterizing the Universe from its initial high energy state through the structure building sequence to the current epoch of accelerated expansion is one of the main issues in modern cosmology.  Our comprehension of early time dynamics and isotropization has long benefited from studies of anisotropic and viscous cosmological models \cite{ref1,ref2,ref3,ref4}.  Other theoretical frameworks, such Chaplygin type fluids \cite{chap01,chap02,chap03,chap04} and modified gravity, offer complementary methods to solve cosmological conundrums and connect early and late time behavior \cite{ref5,ref6,ref7,ref8,ref9}.  Dissipative processes can affect global dynamics and possibly prevent singular behavior in idealized models, as demonstrated by fundamental studies into bulk viscous effects and isotropy \cite{ref10,ref11}.Theoretical models are anchored by observational probes like as direct Hubble parameter determinations, Type Ia supernovae and precise measurements of the cosmic microwave background (CMB).  The Universe's large scale homogeneity and anisotropy have been limited by landmark CMB analyses and associated datasets, which have also placed strict limits on cosmological parameters \cite{ref12,ref13,ref14,ref15,ref16}.  The expansion history and geometry of spacetime are further constrained by independent observations of the Hubble parameter \cite{ref17,ref18}.  Collectively, these findings support both minor adjustments to the concordance model and more drastic substitutes, and they offer the empirical foundation for testing Chaplygin gas models, viscous fluids, and modified gravity situations \cite{ref19,ref20,ref21,ref22}.The most straightforward explanation for late time acceleration is still the cosmological constant $\Lambda$. However, it suffers from conceptual issues, particularly the coincidence problem and fine tuning, which have led to a number of theoretical assessments and proposals \cite{ref23,ref24,ref25,ref26,ref27,ref28}.  Different phenomenologies for cosmic acceleration and structure growth are provided by braneworld setups, tachyonic matter, and dynamical dark energy models (such as quintessence, phantom, and related scalar field constructions) \cite{ref29,ref30,ref31,ref32,ref33,ref34}.  In parallel, geometric pathways to simulate dark energy behavior without utilizing a pure vacuum energy term are shown via holographic and modified gravity frameworks (such as $f(R)$ and Gauss Bonnet inspired structures) \cite{ref35,ref36,ref37,ref38,ref39}.An equation of state that interpolates between dust like and cosmological constant like behavior characterizes the Chaplygin gas family, which includes the original formulation as well as generalized and modified variants. This class of unified dark matter and dark energy models has been thoroughly studied \cite{ref40,ref41,ref42,ref43,ref44,ref45}.  Chaplygin type fluids' thermodynamic characteristics, phase space dynamics, and cosmological implications have been studied in a variety of settings, including brane world and higher dimensional scenarios as well as Gauss–Bonnet extensions \cite{ref46,ref47}.  These studies demonstrate the flexibility of Chaplygin structures as well as the need for phenomenological adjustments to accommodate findings at intermediate redshifts.When modeling realistic cosmic fluids, dissipative processes, and bulk viscosity in particular, are physically well motivated extensions because expansion and perturbation dynamics are changed by viscous stresses that are naturally generated by deviations from local thermodynamic equilibrium \cite{ref48}.  The incorporation of bulk viscous terms into Kaluza Klein frameworks, FRW and anisotropic cosmologies, and different modified gravity settings has demonstrated that viscosity can function as an effective negative pressure and aid in smooth phase transitions or late time acceleration \cite{ref49,ref50,ref51,ref52}.  An attractive unified image that can reduce data tensions while taking into account entropy formation and particle creation processes is produced by combining viscous effects with Chaplygin type equations of state \cite{ref53,ref54,ref55,ref56}.From a statistical and observational standpoint, more accurate datasets necessitate strong inference frameworks in order to investigate multidimensional parameter spaces and measure uncertainty.  This endeavor relies heavily on Markov Chain Monte Carlo (MCMC) techniques, which allow for accurate posterior estimate even in cases when likelihood surfaces are multimodal or non Gaussian \cite{ref57}.  Modified Gauss–Bonnet or $f(R)$ parametrizations have been tested against large scale structure data, probed interacting and holographic extensions, and viscous Chaplygin gas parameters have been constrained in recent work using MCMC and other Bayesian tools \cite{ref58}.  Additionally, growth rate measurements and matter power spectrum studies offer complementary handles on modified gravity and viscous signals \cite{ref58}.Different theoretical frameworks have shaped the study of gravitational and cosmological models.  In order to investigate dark energy--matter interactions, Cold Dark Matter models with a smooth component \cite{ref59} were introduced. An alternate method of avoiding the Big Rip singularity was offered by the $hessence$ dark energy model \cite{ref60}, subsequently rebuilt and examined with supernova data \cite{ref61}. To further comprehend cosmic acceleration, a novel interaction between dark matter and generalized Chaplygin gas was proposed \cite{ref62}. Modified gravity extensions were investigated using $(2+1)$-dimensional cosmological models in $f(R,T)$ \cite{ref63} gravity and further reinforced by advancements in time-scale inequalities mathematics \cite{ref64}. Compact stellar configurations in $f(R,T)$ gravity: an investigation \cite{ref65} and These frameworks were extended by two-fluid cosmological models in scalar tensor theory \cite{ref66}. Additionally, research on higher-dimensional spacetime's weird quark matter \cite{ref67} and Understanding cosmic evolution under modified gravity theories was greatly aided by divergence-free deceleration parameters in Weyl-type $f(Q,T)$ gravity \cite{ref68}.

The study of bulk viscous cosmology with modified Chaplygin gas (MCG) in the FRW spacetime is motivated by the quest to resolve existing observational tensions in cosmology, particularly the discrepancies in the Hubble parameter, by adopting more realistic descriptions of cosmic fluids beyond the perfect-fluid framework \cite{ref69,ref70,ref76,ref77}. Nojiri and Odintsov \cite{ref75} advanced an inhomogeneous equation of state to explain the universe's transition across the phantom split. The MCG model, which naturally unifies dark matter and dark energy through its equation of state, gains further significance when bulk viscosity is included, as it accounts for dissipative processes and entropy production in cosmic evolution\cite{ref71,ref72}.The current study is situated in this framework, driven by the necessity to integrate these themes into a comprehensive model that reflects the dynamics of the universe from its earliest eras to its current accelerated expansion, all the while being consistent with the abundance of observational data\cite{ref73,ref74}. Paul \cite{ref78} newly interrogated bulk viscosity's thermodynamic genesis and associated it to nonequilibrium events in cosmic fluids. By using MCMC statistical methods to limits model parameters, Joshi et al. \cite{ref79} and Mune et al. \cite{ref80} examined anisotropic Bianchi-type cosmologies in the view of modified $f(R,T)$ gravity as a correlate to viscous and thermodynamic models. Together, these study provides reasonable illustration of how changed gravity frameworks and viscous effects can explain the accelerated dynamics of the universe. Sections 2 and 3 deals with the FRW line element and it's solution. Observational Data analysis with Hubble and Pantheon data is given in Section 4. MCMC plots are incorporated into this section with their significance. In Section 5 we define the Hubble and deceleration parameter. Stability analysis is incorporated in section 6 and Concluding remarks in section 7. 

\section{Friedmann Robertson-Walker (FRW) line element}
 
 The following metric describes the Friedmann Robertson-Walker(FRW) universe in four-dimensional space-time.
 \cite{ref13,ref31},
\begin{equation}
ds^{2} = -dt^{2} + a^{2}(t) \left( \frac{dr^{2}}{1 - k r^{2}} + r^{2} d\theta^{2}+r^2\sin^{2}\theta d\phi^{2} \right),
\label{1}
\end{equation}

and the scale factor is denoted by $a(t)$.With $0 \leq \theta \leq \pi$ and $0 \leq \phi < 2\pi$, the $\theta$ and $\phi$ parameters are the standard azimuthal and polar angles of spherical coordinates.Comoving coordinates are the coordinates $(t,r,\theta,\phi)$. Additionally, the space's curvature is indicated by the constant $k$.Only the condition when $k=0$, or flat space, is taken into consideration in this study.The Einstein equation is then provided by

\begin{equation}
R_{ij} - \tfrac{1}{2} g_{ij} R = T_{ij} + g_{ij} \Lambda,
\label{2}
\end{equation}
where $8\pi G = 1$ and $c = 1$ were expected.Additionally, the following connection provides the energy-momentum tensor for the bulk viscous fluid and modified Chaplygin gas \cite{ref5, ref9, ref45, ref3, ref27, ref28}.
\begin{equation}
T_{ij} = (\rho + \bar{p}) u_{i} u_{j} - \bar{p}\, g_{ij},
\label{3}
\end{equation}
If the velocity vector is $u^{i}$ and the energy density is $\rho$,with the normalization condition being $u^{i} u_{j} = -1$.Additionally, the following equations provide the total pressure and appropriate pressure, which incorporate the bulk viscocity coefficient $\xi$ and the Hubble expansion parameter  $H = \dot{a}/a$:

\begin{equation}
\bar{p} = p - 3 \xi H,
\label{4}
\end{equation}

and

\begin{equation}
p = \gamma \rho - \frac{A}{\rho^{\alpha}},
\label{5}
\end{equation}
With $0 < \alpha \leq 1$ and $A > 0$.One of the most crucial numbers for characterizing the characteristics of dark energy theories is the equation of state parameter $\gamma$.Bulk viscocity is clearly represented by the parameter $\zeta$,whereas the influence of the Chaplygin gas is represented by $A$.The dynamics of FRW cosmology with modified Chaplygin gas as the matter source were already developed in \cite{ref16}.The eigenvalues of the linearized Jacobi matrix for the exceptional situation $\alpha = 0.6$ were then evaluated in order to study the nature of the critical points.In this paper, we extend the earliar work \cite{ref16} to incorporate the bulk viscous coefficient and analyze the exceptional situation $\alpha = 0.5$.

In that case the Friedmann equations are given by
\begin{equation}
\left( \frac{\dot{a}}{a} \right)^{2} = \frac{\rho}{3},
\label{6}
\end{equation}
Equation \eqref{6} can be written as
\begin{equation}
H^2=\frac{\rho}{3}
\label{7}
\end{equation}
and

\begin{equation}
2 \frac{\ddot{a}}{a} + \left( \frac{\dot{a}}{a} \right)^{2} = -\bar{p},
\label{8}
\end{equation}

where the derivative with regard to the cosmic time $t$ is indicated by the dot.The following is the energy momentum conservation law:

\begin{equation}
\dot{\rho} + 3 \frac{\dot{a}}{a} (\rho + \bar{p}) = 0.
\label{9}
\end{equation}

\begin{equation}
\dot{\rho} + 3 H (\rho + \bar{p}) = 0.
\label{10}
\end{equation}
Using the above equations, we attempt to determine the time-dependent density in the next section.

\section{Solution of a Friedmann Robertson-Walker(FRW) universe}

Using the conservation relation Equation \eqref{9} and Equations \eqref{4},\eqref{5} and \eqref{6}, we have
\begin{equation}
\dot{\rho} + \sqrt{3} (\gamma + 1) \rho^{\tfrac{3}{2}} - 3 \xi \rho - \sqrt{3} A = 0.
\label{11}
\end{equation}
Case (I) : $\xi = 0$, then one can extract the energy density 
as a function of the scale factor \cite{ref16}.

\begin{equation}
\rho(a) = \left[ \frac{1}{\gamma+1} \left( A + \frac{d}{\sqrt{a^{9(\gamma+1)}}} \right) \right]^{\tfrac{2}{3}},
\label{12}
\end{equation}

where $d$ is an integration constant. 

Put the value of Equation \eqref{12} in Equation \eqref{7},we get

\begin{equation}
H(a) = \frac{1}{\sqrt{3}}\left[ \frac{1}{\gamma+1} \left( A + \frac{d}{\sqrt{a^{9(\gamma+1)}}} \right) \right]^{\tfrac{1}{3}},
\label{13}
\end{equation}

as  
    $$a=(1+z)^{-1}$$

Equation \eqref{12} can be written as

\begin{equation}
\rho(z) = \left[ \frac{1}{\gamma+1} \left( A + {d}(1+z)^{3\sqrt{\gamma+1}} \right) \right]^{\tfrac{2}{3}},
\label{14}
\end{equation}

Put Equation \eqref{14} in Equation \eqref{7}, then we get,

\begin{equation}
H(z) = \frac{1}{\sqrt{3}}\left[ \frac{1}{\gamma+1} \left( A + {d}(1+z)^{3\sqrt{\gamma+1}} \right) \right]^{\tfrac{1}{3}},
\label{15}
\end{equation}

\begin{equation}
H(z) = H_0\left[  \frac{\left( A + {d}(1+z)^{3\sqrt{\gamma+1}} \right)}{A+d} \right]^{\tfrac{1}{3}},
\label{16}
\end{equation}

Case(II) : $\xi\neq0$ (Constant)

In this case, we adhere to the specific $\rho$ from previously provided by Saadat and Pourhassan.

\begin{equation}
\rho = \frac{B}{t^{2}} + \frac{F}{t} + h t + E e^{b t},
\label{17}
\end{equation}

where the constants $B$, $F$, $h$, $E$, and $b$ should be determined. 
Substituting relation \eqref{17} into Equation \eqref{11}, we get

$$0=\frac{d}{dt}[\frac{B}{t^2}+\frac{F}{t}+ht+Ee^{bt}]+\sqrt{3}(\gamma+1)[\frac{B}{t^2}+\frac{F}{t}+ht+Ee^{bt}]^{\frac{3}{2}}-3\xi(\frac{B}{t^2}+\frac{F}{t}+ht+Ee^{bt})-\sqrt{3}A$$

By comparing both sides, gives us the following coefficients:

\begin{equation}
h = \sqrt{3} A,
\label{18}
\end{equation}

\begin{equation}
B = \frac{4}{3} {(\gamma + 1)^{-2}},
\label{19}
\end{equation}

\begin{equation}
F = 2 \xi (\gamma + 1)^{-2},
\label{20}
\end{equation}

\begin{equation}
E = \frac{(\gamma + 1)^{2}}{4} \left[ \frac{8 \sqrt{3} \xi^{2}}{(\gamma + 1)^{3}}
      - \frac{3 (\gamma + 1)^{4}}{16 \xi^{2}} \right],
      \label{21}
\end{equation}

\begin{equation}
b = \frac{\xi \big[ \sqrt{3}\xi (\gamma+1)(A(\gamma+1) - \tfrac{9}{2}\xi^{3}) 
   + \tfrac{27}{32} (\gamma + \tfrac{1}{8}) 
   + \tfrac{9}{16} \xi^{4} + \mathcal{O}(\gamma^{n}) \big]}
   {8(\gamma+1)\big( \sqrt{3}\xi^{4} - \tfrac{3}{128}(\gamma+1)^{7} \big)},
   \label{22}
\end{equation}
where,

\begin{equation}
\mathcal{O}(\gamma^n) \equiv 
\frac{189}{64}\gamma^2 + 
\frac{189}{32}\gamma^3 + 
\frac{945}{128}\gamma^4 + 
\frac{189}{32}\gamma^5 + 
\frac{189}{64}\gamma^6 
+ \frac{27}{32}\gamma^7 + 
\frac{27}{256}\gamma^8 .
\label{23}
\end{equation}

After putting all the values of the arbitrary constant in Equation \eqref{17}, we get

\begin{equation}
    \rho=\frac{4}{3(\gamma+1)^{2}t^2}+\frac{2\xi}{(\gamma+1)^{2}t}+\sqrt{3}At+Y_1e^{Y_2}
    \label{24}
\end{equation}

where we define,

$$Y_1=\frac{(\gamma+1)^2}{4}[\frac{8\sqrt{3}\xi^2}{(\gamma+1)^3}-\frac{3(\gamma+1)^4}{16\xi^2}]$$

$$Y_2=Y_3[8(\gamma+1)(\sqrt{3}\xi^4-\frac{3{(\gamma+1)^7}}{128})]^{-1}t$$

$$Y_3=\xi(Y_4+\frac{9\xi^4}{16}+\mathcal{O}(\gamma^n))$$

$$Y_4=Y_5+\frac{27}{32}(\gamma+\frac{1}{8})$$

$$Y_5=\sqrt{3}\xi(\gamma+1)(A(\gamma+1)-\frac{9}{2}\xi^3)$$

from Equation \eqref{7}, the Hubble parameter can be written as 
\begin{equation}
    H=\sqrt\frac{\rho}{3}
    \label{25}
\end{equation}

The equation for red shift is $1+z=\frac{a_0}{a}$,which also requires the normalized scale factor $a_0=1$. Following the establishment of the $t-z$ relationship, $t(z)=\frac{1}{n\alpha}log(1+(1+z)^{-n})$ is obtained.
To formulate $\rho(z)$, let consider
$$Y(z)=1+(1+z)^{-n}$$

$$t(z)=\frac{1}{n\alpha}log(Y(z))$$

$$\frac{1}{t}=\frac{n\alpha}{logY}$$

that implies, $$\frac{1}{t^2}=\frac{n^2\alpha^2}{(logY)^2}$$

for finding the values of $\rho(z)$ we will find termwise

$$\frac{4}{3(\gamma+1)^2t^2}=\frac{4n^{2}\alpha^2}{3(\gamma+1)^{2}(logY)^2}$$

$$\frac{2\xi}{(\gamma+1)^{2}t}=\frac{2\xi n\alpha}{(\gamma+1)^{2}(logY)}$$

$$\sqrt{3}At=\frac{\sqrt{3}A}{n\alpha}logY$$

as,
$$Y_2=Y_3\left[8(\gamma+1)(\sqrt{3}\xi^4-\frac{3}{128}(\gamma+1)^7)\right]^{-1}\frac{logY}{n\alpha}$$

Consider,
$$M=Y_3\left[8(\gamma+1)\left(\sqrt{3}\xi^4-\frac{3}{128}(\gamma+1)^7\right)\right]^{-1}$$

therefore, $$Y_2=\frac{MlogY}{n\alpha}$$

$$e^{Y_2}=Y^{\frac{M}{n\alpha}}$$

from Equation \eqref{23},
\begin{equation}
    \rho(z)= \frac{4n^{2}\alpha^2}{3(\gamma+1)^{2}(logY)^2}+\frac{2\xi n\alpha}{(\gamma+1)^{2}(logY)^2}+\frac{\sqrt3A}{n\alpha}logY+\left(\frac{(\gamma+1)^2}{4}\left[\frac{8\sqrt{3}\xi^2}{(\gamma+1)^3}-\frac{3(\gamma+1)^4}{16\xi^2}\right]\right)Y^{\frac{M}{n\alpha}}
    \label{26}
\end{equation}

as we have,

$$H(a)=\frac{1}{\sqrt{3}}\rho^{\frac{1}{2}}$$

$$H(z)=\frac{1}{\sqrt{3}}[\rho(z)]^{\frac{1}{2}}$$

from Equation \eqref{25}, we get

\begin{equation}
    H(z)= \frac{1}{\sqrt{3}}\left[\frac{4n^{2}\alpha^2}{3(\gamma+1)^{2}(logY)^2}+\frac{2\xi n\alpha}{(\gamma+1)^{2}(logY)^2}+\frac{\sqrt3A}{n\alpha}logY+\left(\frac{(\gamma+1)^2}{4}\left[\frac{8\sqrt{3}\xi^2}{(\gamma+1)^3}-\frac{3(\gamma+1)^4}{16\xi^2}\right]\right)Y^{\frac{M}{n\alpha}}\right]^{\frac{1}{2}}
    \label{27}
\end{equation}

\begin{equation}
    H(z)= H_0\left[\frac{\frac{4n^{2}\alpha^2}{3(\gamma+1)^{2}(logY)^2}+\frac{2\xi n\alpha}{(\gamma+1)^{2}(logY)^2}+\frac{\sqrt3A}{n\alpha}logY+\left(\frac{(\gamma+1)^2}{4}\left[\frac{8\sqrt{3}\xi^2}{(\gamma+1)^3}-\frac{3(\gamma+1)^4}{16\xi^2}\right]\right)Y^{\frac{M}{n\alpha}}}{\frac{4n^{2}\alpha^2}{3(\gamma+1)^{2}(log2)^2}+\frac{2\xi n\alpha}{(\gamma+1)^{2}(log2)^2}+\frac{\sqrt3A}{n\alpha}log2+\left(\frac{(\gamma+1)^2}{4}\left[\frac{8\sqrt{3}\xi^2}{(\gamma+1)^3}-\frac{3(\gamma+1)^4}{16\xi^2}\right]\right)2^{\frac{M}{n\alpha}}}\right]^{\frac{1}{2}}
    \label{28}
\end{equation}
where,
$$Y=1+(1+z)^{-n}$$

Case (III):If we neglect both bulk viscosity and presence of Chaplygin gas, i.e
$A=0$ and $\xi=0$, our Equation \eqref{11} becomes
\begin{equation}
    \dot{\rho} + \sqrt{3} (\gamma + 1) \rho^{\tfrac{3}{2}}=0
    \label{29}
\end{equation}

After solving Equation \eqref{29}, we get

\begin{equation}
\rho^{\frac{1}{2}}=\frac{2}{\sqrt{3}(\gamma+1)t}
\label{30}
\end{equation}

therefore Equation \eqref{30} becomes
\begin{equation}
\rho = \frac{4}{3(\gamma+1)^2 t^2}.
\label{31}
\end{equation}

from Equation \eqref{7}, we get

\begin{equation}
H = \frac{2}{3(\gamma+1) t}.
\label{32}
\end{equation}

Put the value of t in terms of z, we get
\begin{equation}
H(z) = \frac{2n\alpha}{3(\gamma+1) log(1+(1+z)^{-n})}.
\label{33}
\end{equation}

Equation \eqref{33} can be written as
\begin{equation}
    H(z)=H_0 \left[\frac{log2}{log(1+(1+z)^{-n})}\right]
    \label{34}
\end{equation}

The findings of earlier studies \cite{ref29, ref16}, where $\rho \propto t^{-2}$ was established, are consistent with this.  On the other hand, $b < 0$ for a large bulk viscosity coefficient, which leads to the result $\rho \propto \xi / t$.  Additionally, a constant negative energy density is obtained for the situation of infinitesimal $\xi$.Examining the density's late-time activity is intriguing. 
In such instance, $\rho \sim E e^{bt}$ can be stated since the final term of Equation \eqref{17} is dominant.  In the general situation, the energy density is a decreasing function of time, according to Equation \eqref{17} with the coefficients in Equations \eqref{18}-\eqref{22}. 

\section{Observational data analysis}

Let's use recent observational datasets, specifically the observed Hubble parameter measurements (OHD) and Pantheon data, to test the model's feasibility. These datasets are especially helpful for verifying that the model is consistent with large-scale structure observations and offer complementary restrictions on the cosmic expansion history.

\subsection{Hubble Data}

The differential age (cosmic chronometer) methodology and Pantheon compilation method are two popular methods in cosmology for estimating the Hubble parameter $H(z)$ at a specific redshift \cite{ref48,ref49}. 

 We use 30 observational measurements of $H(z)$ in the redshift range of $0.07 < z < 1.965$ in this work.  As stated in 
 
\begin{equation}
\chi^{2}_{H} = \sum_{i=1}^{30} \frac{\left( H_{\mathrm{th}} - H_{\mathrm{obs}} \right)^{2}}{\sigma_{i}^{2}}
\label{35}
\end{equation}

where $H_{\mathrm{th}}$, $H_{\mathrm{obs}}$, and $\sigma_{H(z_{i})}$ represent the standard error, the observed value, and the theoretical prediction of the Hubble parameter, respectively, for the 30 data points taken into consideration.

\subsection{Pantheon data}

For the Pantheon dataset, the model parameters on current investigation can be fitted by comparing the observed $\mu_i^{\text{Obs}}$ 
with the theoretical $\mu_i^{\text{Th}}$ value of the distance moduli as follows:

$$\mu = m - M_{B} = 5\log_{10}(D_L) + \mu_0$$

where $M$ and $m$ represents the absolute and apparent magnitudes respectively and $\mu_0 = 5\log\left( H_0^{-1} / \text{Mpc} \right) + 25$ is the nuisance parameter that has been depreciate\cite{ref53}.
Now, the luminosity distance can be defined as

$$D_L(z) = \frac{c}{H_0} (1+z) \int_{0}^{z} \frac{dz^*}{E(z^*)}$$

The $\chi^2$ function of the (SNIa) measurements is given by

$$\chi^2_{SN}(\phi^\nu_s) = \mu_s C^{-1}_{s,\text{cov}} \mu_s^{T}$$

where,
$$ \mu_s = \{\mu_1 - \mu_{\text{th}}(z_1, \phi_s^{\nu}), \ldots, \mu_N - \mu_{\text{th}}(z_N, \phi_s^{\nu})\} $$

\subsection{Markov chain Monte Carlo}
\subsubsection{Case(I) : Model 1}

\begin{figure}[H]
    \centering
    \includegraphics[width=0.6\textwidth]{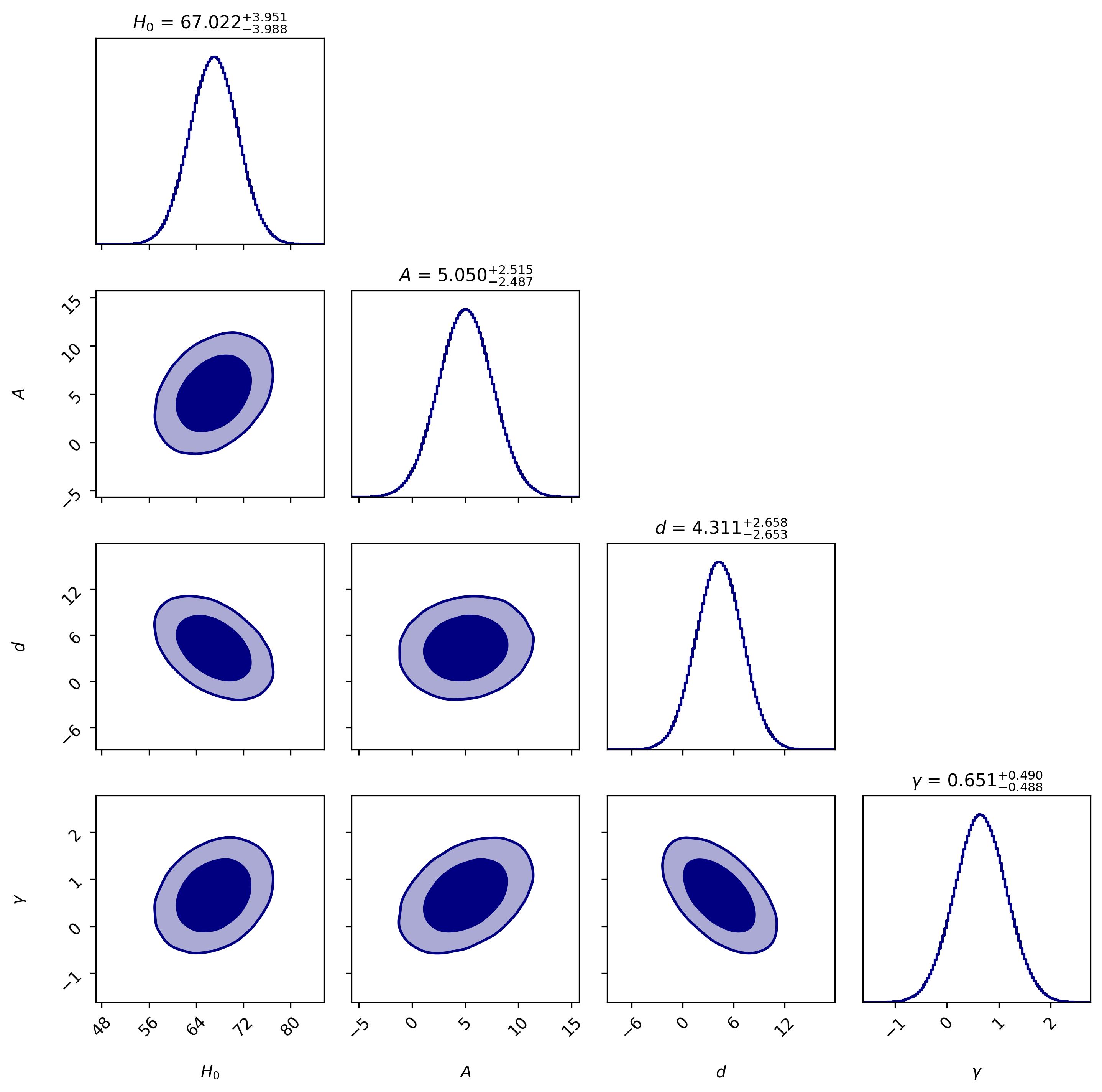} 
    \caption{Model(1) Confidence contours and marginalized posterior distributions for the parameters $H_0$, $A$, $d$ and $\gamma$ obtained from the Hubble (OHD) dataset.}
   \label{figure:1}
\end{figure}

Applying a Markov Chain Monte Carlo (MCMC) investigation, the combined posterior distributions of the model parameters $H_0$, $A$, $d$ and $\gamma$ are shown in the corner plot in Figure~\ref{figure:1}. The one dimensional marginalization probability allocations for each parameter are shown in diagonal panels simultaneously with their $68\%$ confidence interval. The plot  shows two confidence regions contours corresponding to $1\sigma$ ($68\%$ confidence level) and $2\sigma$ ($95\%$ confidence level), where inner navy blue area shows the $1\sigma$ uncertainity and outer light navy blue area shows $2\sigma$ uncertainity. The model selection statistics denotes $\mathrm{AIC}_{\text{model}} = 19.59$, $\mathrm{BIC}_{\text{model}} = 25.19$ and $\mathrm{DIC}_{\text{model}} = 13.22$. The best fit best-fit chi-square value is $\chi^{2} = 11.587$, resultant in a reduced chi-square $\chi^{2}_{\text{red}} = 0.446$. This shows that the model and observational data have great consensus. In general, the findings indicates that the model gives a true and consonant chronology of the cosmological data and that decided parameter behaves fine.

\begin{figure}[H]
    \centering
    \includegraphics[width=0.6\textwidth]{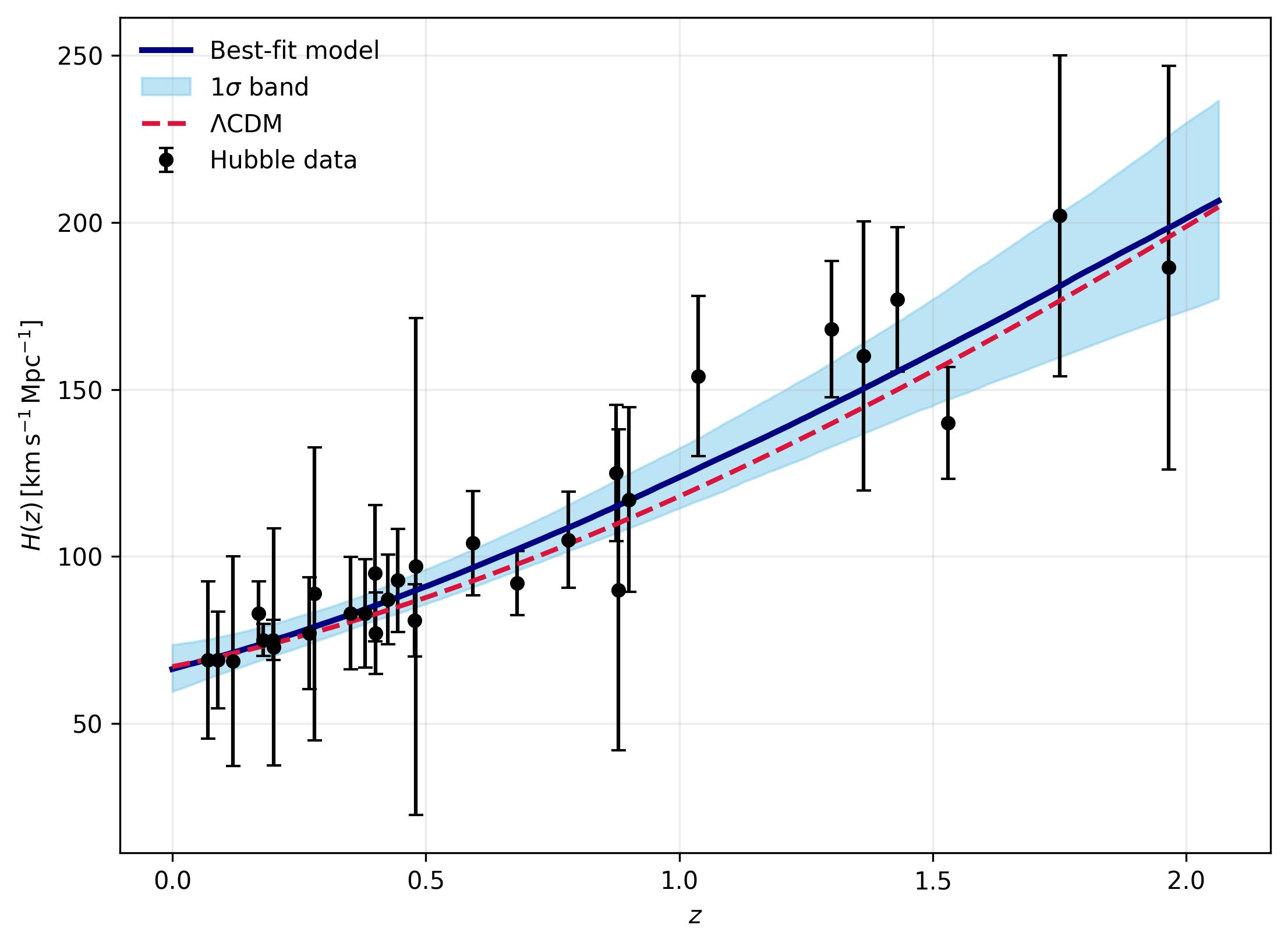} 
    \caption{Model(1) Hubble data Error graph compare with $\Lambda\ CDM$ model}
     \label{figure:2}
\end{figure}

Direct comparison with empirical Hubble parameter data can be used to evaluate the observational validation of model (1). Figure~\ref{figure:2} presents our Model's predictions , displayed next to the traditional $\Lambda$CDM model (red dashed curve). The graphic includes 30 observational Hubble data points from the redshift interval $0.07 < z < 2.0$, each of which is shown with an accompanying error bar. Our model's MAP best-fit accurately captures the overall pattern of the observational data, especially at low to intermediate redshift values. The model curve stays well within the observational uncertainties, despite minor variations that show up at higher redshifts, demonstrating the model's resilience to actual data.

\begin{figure}[H]
    \centering
    \includegraphics[width=0.5\textwidth]{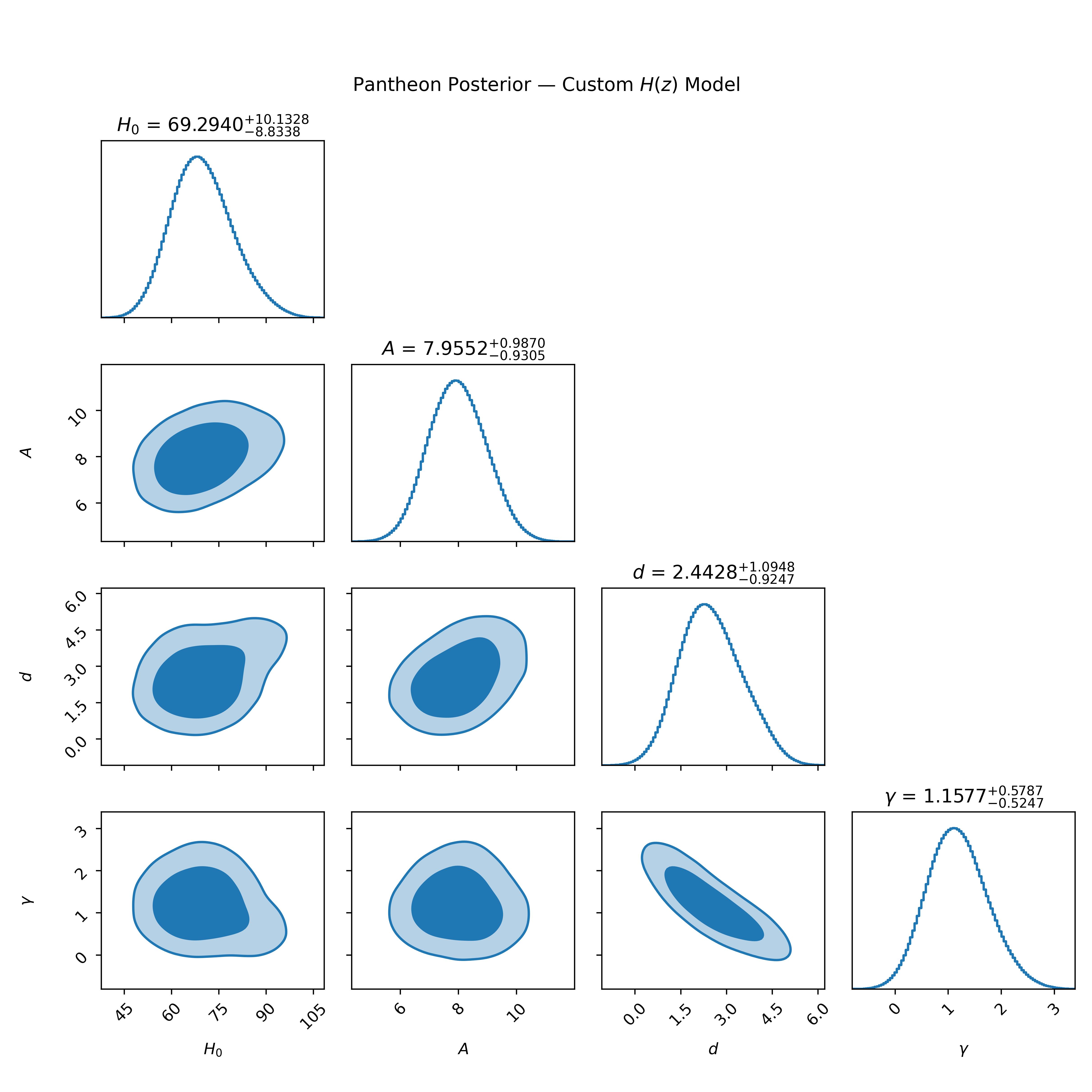} 
    \caption{Model(1) MCMC-based posterior distributions for parameters $H_0$, $A$, $d$ and $\gamma$ using the Pantheon Type Ia Supernova dataset.}
     \label{figure:3}
\end{figure}

The Pantheon data was used to create contour maps  for four parameters ($H_0$, $A$, $d$ , $\gamma$) which is given in Figure~\ref{figure:3}.The lighter contours area show the 95\% levels, whereas the darker areas provide the 68\% confidence intervals.  Tight constrains on the parameters are shown by the well defined and narrow peaks.  The model selection statistics denotes $\mathrm{AIC}_{\text{model}} = 863.481$, $\mathrm{BIC}_{\text{model}} = 883.296$ and $\mathrm{DIC}_{\text{model}} = 857.401$. The best fit best-fit chi-square value is $\chi^{2} = 855.481$, resultant in a reduced chi-square $\chi^{2}_{\text{red}} = 0.820$. This shows that the model and observational data have great concurrence. In general, the findings indicates that the model gives a true and consonant chronology of the cosmological data and that decided parameter behaves fine.

\begin{figure}[H]
    \centering
    \includegraphics[width=0.6\textwidth]{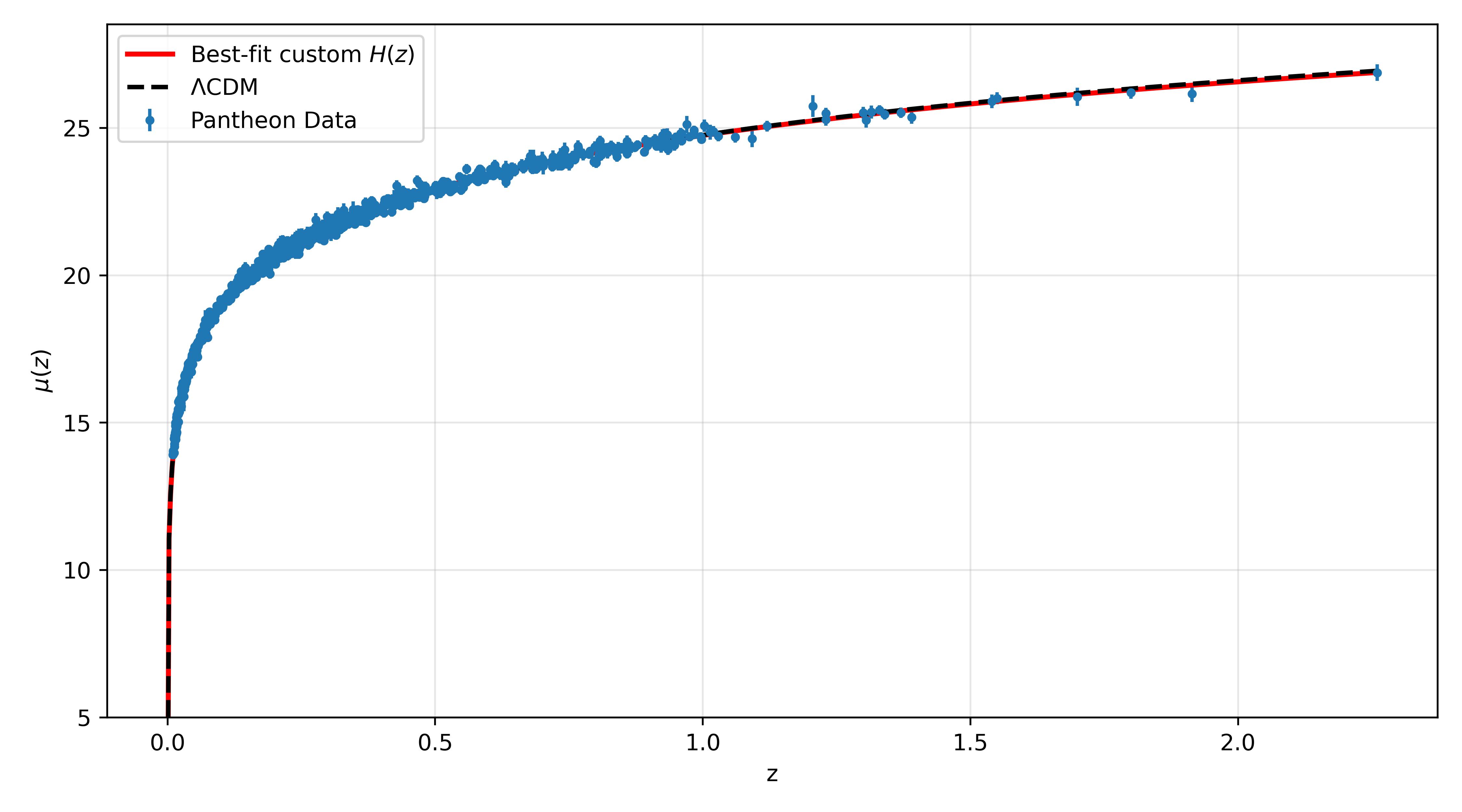} 
    \caption{Model(1) Pantheon data Error graph compare with $\Lambda\ CDM$ model}
     \label{figure:4}
\end{figure}

The Figure~\ref{figure:4} compares the observational data and Pantheon Type la Supernovae ( blue points with error bars ) with the theoretical distance modulus $\mu(z)$ read by the rebuilt cosmological model ( Solid red line ). For visual analogy, the average $\Lambda$ CDM model is described by the black dashed curve, which has been moved to match with data. Above the whole redshift range ($0 < z < 2.5$) , the rebuilt model and the observational data overlap nearly, indicating a good match and high degree of consistency with observational measurements, displaying that the indicated model successfully imitates the observed cosmic acceleration while keeping compliancy with the mainstream cosmological framework.

\begin{figure}[H]
    \centering
    \includegraphics[width=0.5\textwidth]{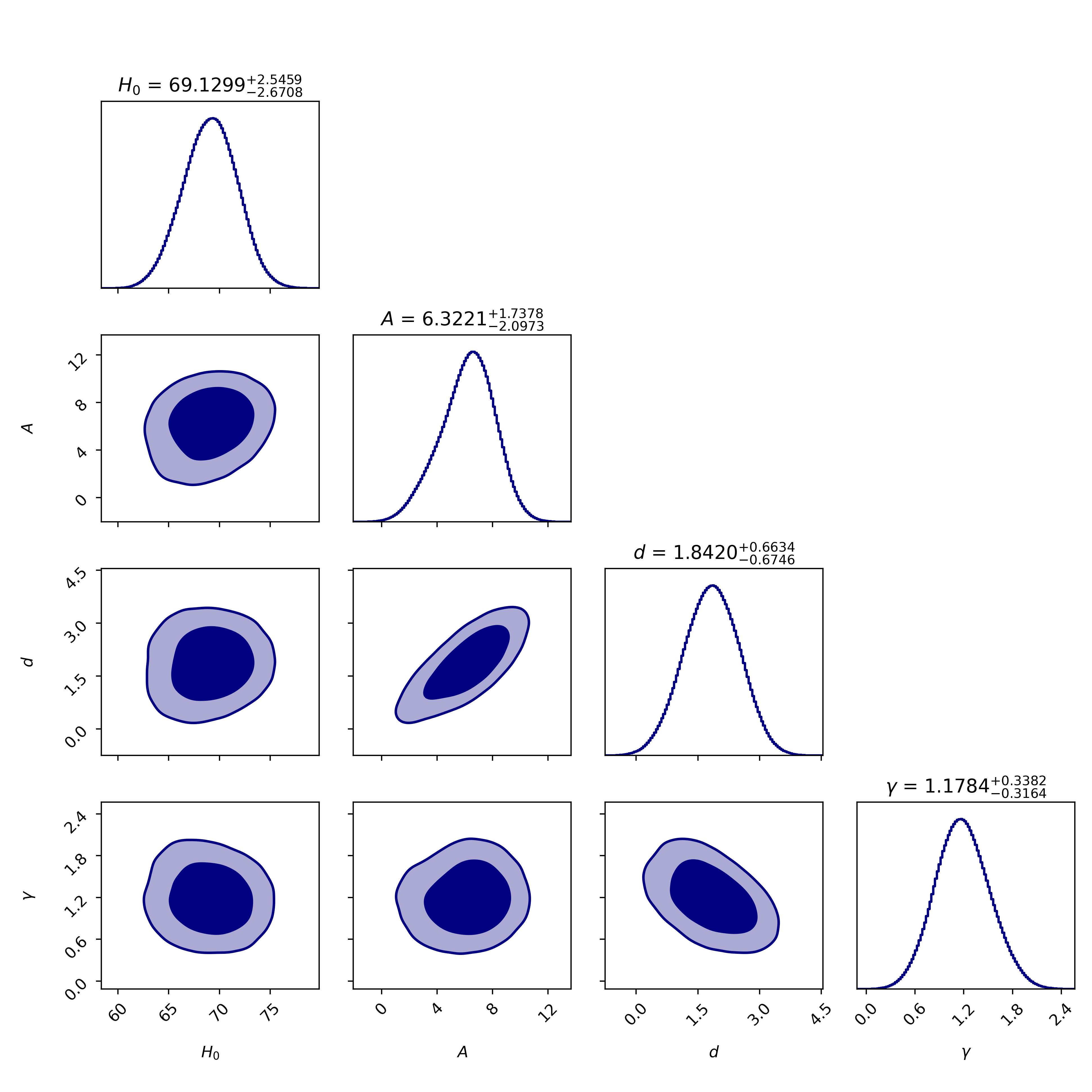} 
    \caption{Model(1) Combined parameter constraints from joint Hubble (OHD) and  Pantheon Type Ia Supernova datasets.}
     \label{figure:5}
\end{figure}

The (Hubble+Pantheon) data was used to create contour maps for four parameters ($H_0$, $A$, $d$ , $\gamma$) which is given in Figure~\ref{figure:5} .The lighter contours area show the 95\% levels, whereas the darker areas provide the 68\% confidence intervals.  Tight constrains on the parameters are shown by the well defined and narrow peaks.  These parameter fits, which are confirmed by the AIC and BIC values.The model selection statistics denotes $\mathrm{AIC}_{\text{model}} = 608.93$, $\mathrm{BIC}_{\text{model}} = 628.86$ and $\mathrm{DIC}_{\text{model}} = 604.96$. The best fit best-fit chi-square value is $\chi^{2} = 600.93$, resultant in a reduced chi-square $\chi^{2}_{\text{red}} = 0.561$,show that the four-parameter model outperforms the  Elementary constant Hubble model in terms of statistically describing the expansion of the universe.

\setlength{\extrarowheight}{2.5 pt}
\begin{table}[H]
\centering

\begin{tabular}{|l|l|l|}
\hline
\textbf{Data used} & \textbf{Parameters} & \textbf{Best fit values} \\ \hline

\multirow{4}{*}{Hubble} 
 & $H_0$ & $67.022^{+3.951}_{-3.988}$ \\ \cline{2-3}
 & $A$ & $5.050^{+2.515}_{-2.487}$ \\ \cline{2-3}
 & $d$ & $4.311^{+2.658}_{-2.653}$ \\ \cline{2-3}
 & $\gamma$ & $0.651^{+0.490}_{-0.488}$ \\ \hline

\multirow{4}{*}{Pantheon} 
 & $H_0$ & $69.2940^{+10.1328}_{-8.8338}$ \\ \cline{2-3}
 & $A$ & $7.9552^{+0.9870}_{-0.9305}$ \\ \cline{2-3}
 & $d$ & $2.4428^{+1.0948}_{-0.9247}$ \\ \cline{2-3}
 & $\gamma$ & $1.1577^{+0.5787}_{-0.5247}$ \\ \hline

 \multirow{4}{*}{Hubble+Pantheon} 
 & $H_0$ & $69.1299^{+2.5459}_{-2.6708}$ \\ \cline{2-3}
 & $A$ & $6.3221^{+1.7378}_{-2.0973}$ \\ \cline{2-3}
 & $d$ & $1.8420^{+0.6634}_{-0.6746}$ \\ \cline{2-3}
 & $\gamma$ & $1.1784^{+0.3382}_{-0.3164}$ \\ \hline

\end{tabular}
\caption{Model (1) fits to cosmological data, showing dataset, parameter sets, and best-fit values.}
\label{tab:fit_stats1}
\end{table}

\begin{table}[H]
\centering
\resizebox{\textwidth}{!}{

\begin{tabular}{|c|c|c|c|c|c|c|c|c|}
\hline
\textbf{Data used} & $\chi^2$ & \textbf{Reduced} $\chi^2$ & \textbf{AIC} & \textbf{BIC} & \textbf{DIC} 
& $\boldsymbol{\Delta\mathrm{AIC}}$ 
& $\boldsymbol{\Delta\mathrm{BIC}}$ 
& $\boldsymbol{\Delta\mathrm{DIC}}$ 
\\
\hline
 Hubble&11.587  &0.446 &19.59& 25.19& 13.22& 6.29& 10.49& 1.92 \\ 
\hline
 Pantheon&855.481&0.820& 863.481&883.296&857.401& 2.913&17.774& -1.167  \\ 
\hline
 Hubble+Pantheon&600.93&0.561&608.93& 628.86&604.96&0.92&10.88&0.95\\ 
 \hline
\end{tabular}}
\caption{Statistical results for different datasets for Model(1).}
\label{tab:fit_stats2}
\end{table}

\subsubsection{Case(II) : Model 2}
\begin{figure}[H]
    \centering
    \includegraphics[width=0.6\textwidth]{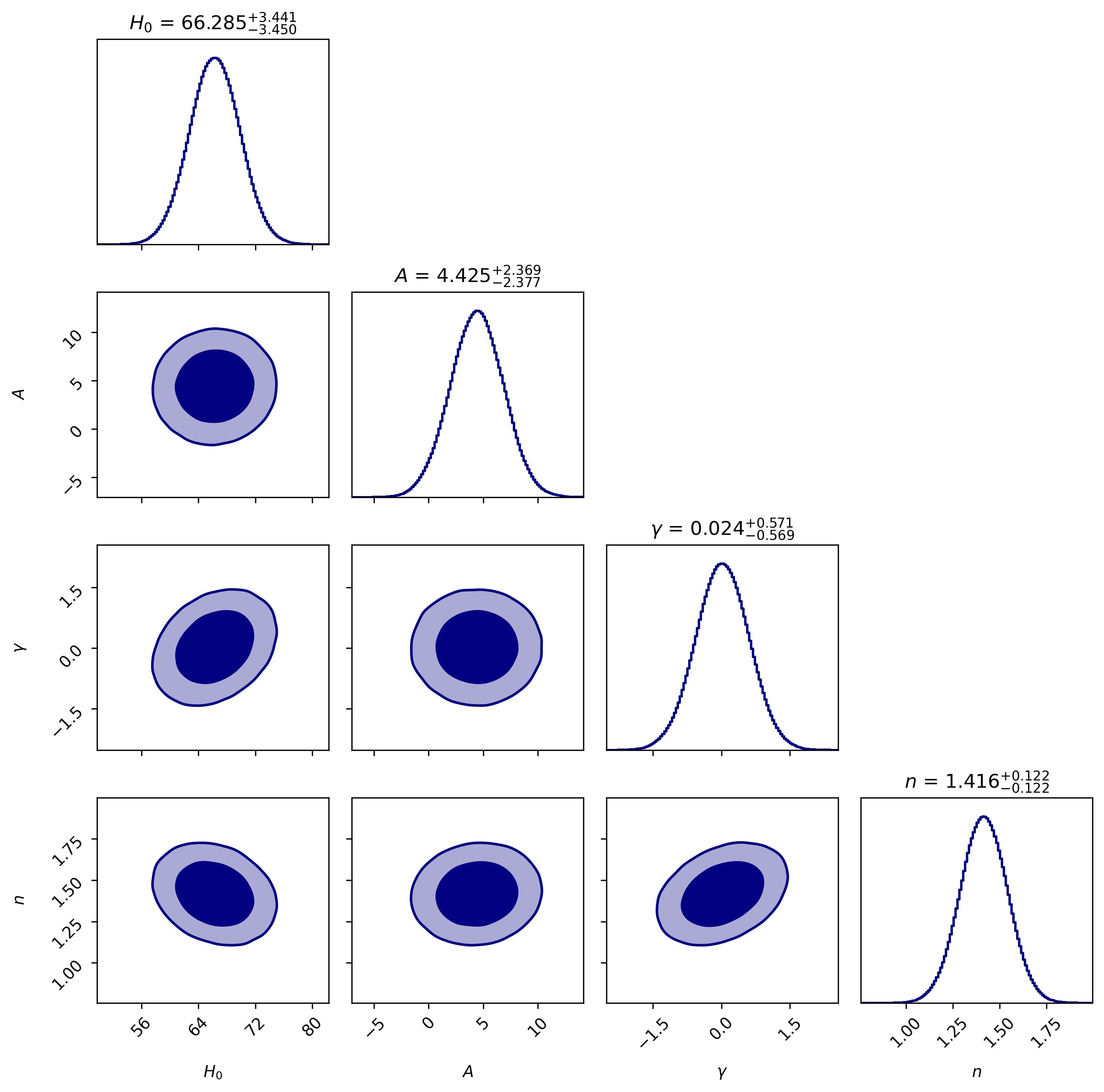} 
    \caption{Model(2) Confidence contours and marginalized posterior distributions for the parameters $H_0$, $A$, $\gamma$ and $n$ obtained from the Hubble (OHD) dataset}
     \label{figure:6}
\end{figure}

The Hubble data was used to create contour maps for four parameters ($H_0$, $A$,$\gamma$, $n$) which is given in Figure~\ref{figure:6}.The lighter contours area show the 95\% levels, whereas the darker areas provide the 68\% confidence intervals.  Tight constrains on the parameters are shown by the well defined and narrow peaks. The model selection statistics denotes $\mathrm{AIC}_{\text{model}} = 20.03$, $\mathrm{BIC}_{\text{model}} = 25.64$ and $\mathrm{DIC}_{\text{model}} = 12.56$. The best fit best-fit chi-square value is $\chi^{2} = 12.033$, resultant in a reduced chi-square $\chi^{2}_{\text{red}} = 0.463$.This shows that the model and observational data have great consensus.

\begin{figure}[H]
    \centering
    \includegraphics[width=0.6\textwidth]{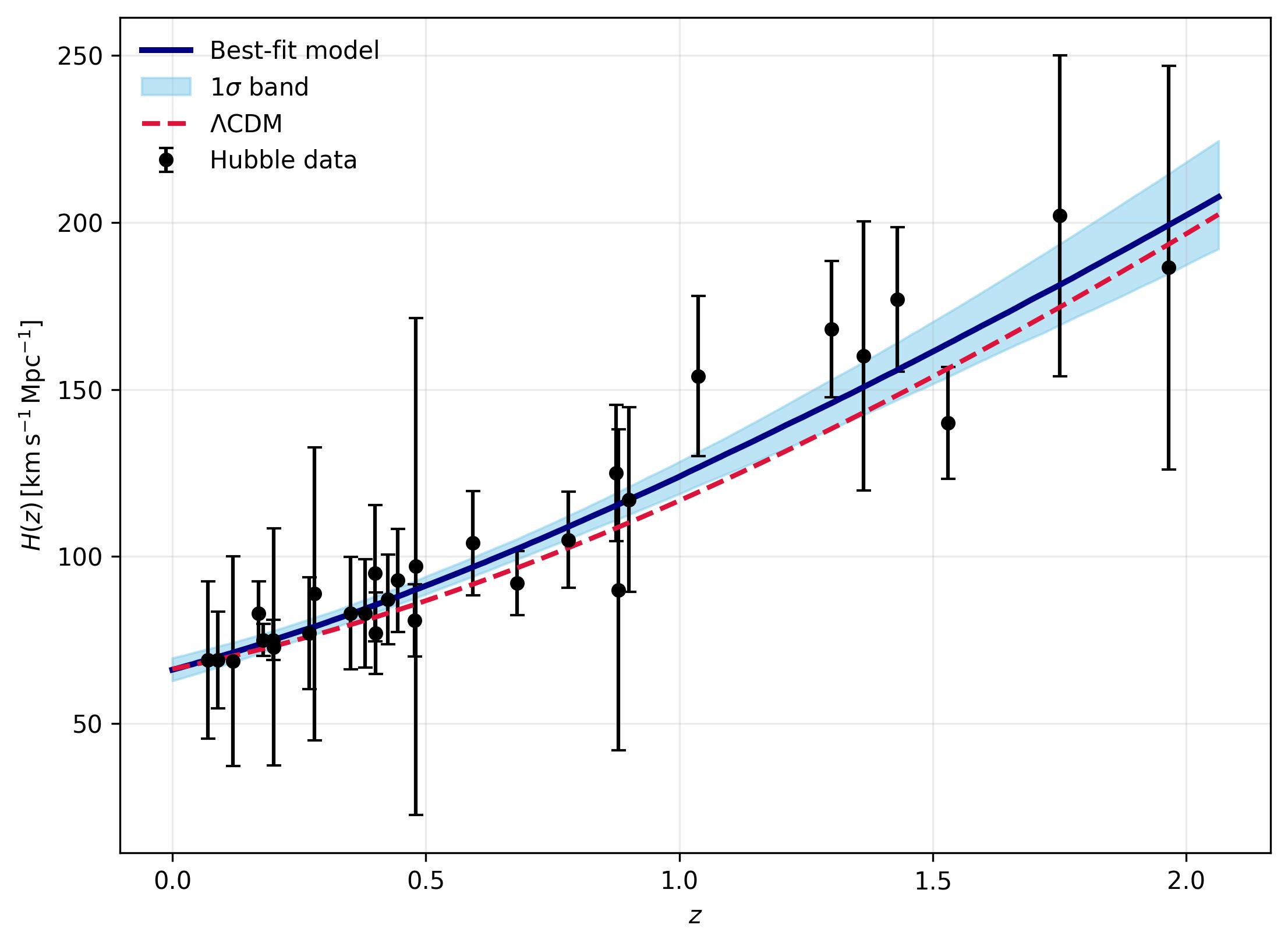} 
    \caption{Model(2) Hubble data Error graph compare with $\Lambda\ CDM$ model}
     \label{figure:7}
\end{figure}

Direct comparison with empirical Hubble parameter data can be used to evaluate the observational validation of model (2). Figure~\ref{figure:7} presents our Model's predictions , displayed next to the traditional $\Lambda$CDM model (red dashed curve). The graphic includes 30 observational Hubble data points from the redshift interval $0.07 < z < 2.0$, each of which is shown with an accompanying error bar. Our model's MAP best-fit accurately captures the overall pattern of the observational data, especially at low to intermediate redshift values. The model curve stays well within the observational uncertainties, despite minor variations that show up at higher redshifts, demonstrating the model's resilience to actual data.

\begin{figure}[H]
    \centering
    \includegraphics[width=0.5\textwidth]{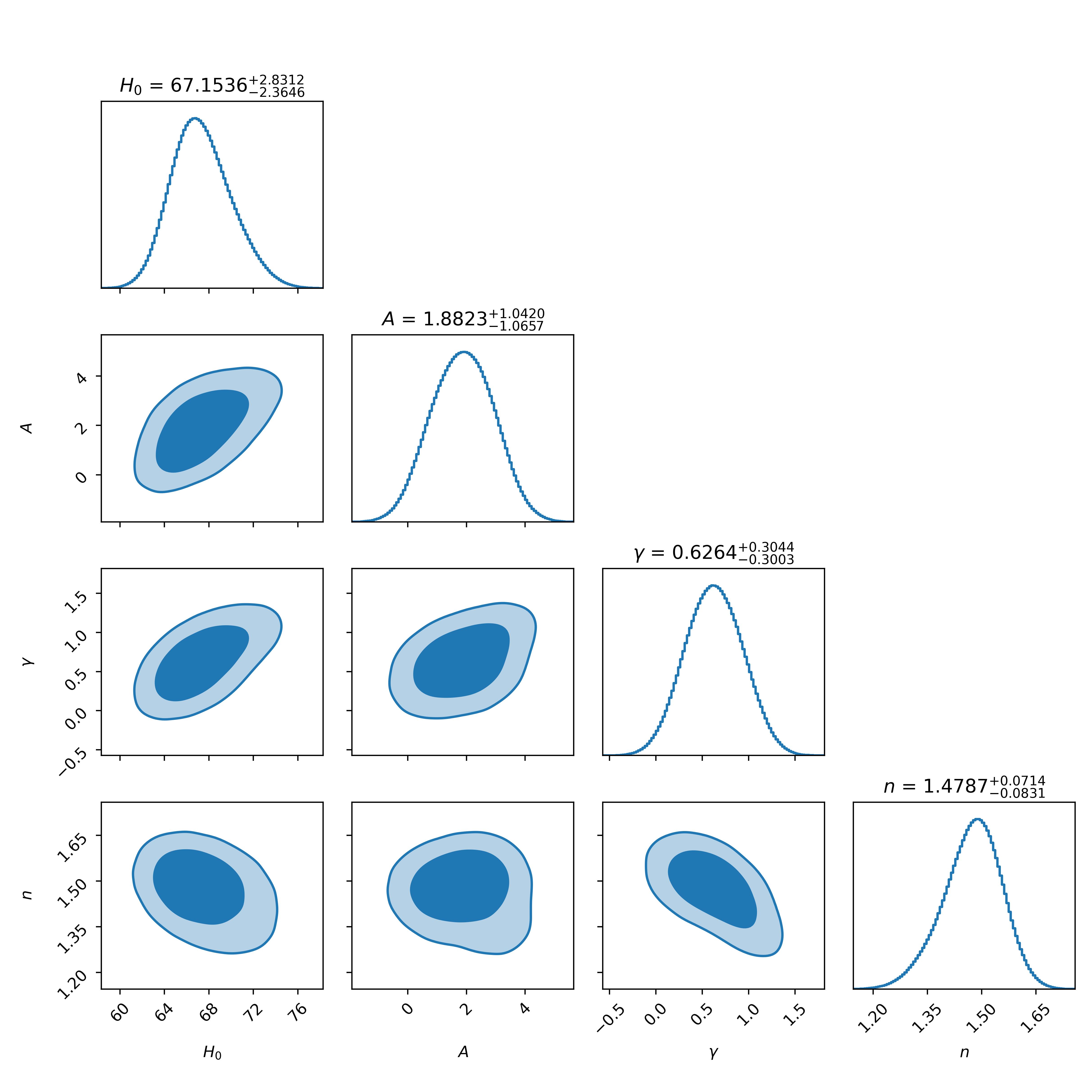} 
    \caption{Model(2) MCMC-based posterior distributions for parameters $H_0$, $A$, $\gamma$ and $n$ using the Pantheon Type Ia Supernova dataset.}
     \label{figure:8}
\end{figure}

The Pantheon data was used to create contour maps for four parameters ($H_0$, $A$,$\gamma$, $n$) which is given in Figure~\ref{figure:8}.The lighter contours area show the 95\% levels, whereas the darker areas provide the 68\% confidence intervals.  Tight constrains on the parameters are shown by the well defined and narrow peaks.  The model selection statistics denotes $\mathrm{AIC}_{\text{model}} = 867.880$, $\mathrm{BIC}_{\text{model}} = 887.695$ and $\mathrm{DIC}_{\text{model}} = 857.162$. The best fit best-fit chi-square value is $\chi^{2} = 859.880$, resultant in a reduced chi-square $\chi^{2}_{\text{red}} = 0.824$.This shows that the model and observational data have great consensus. In general, the findings indicates that the model gives a true and consonant chronology of the cosmological data and that decided parameter behaves fine.

\begin{figure}[H]
    \centering
    \includegraphics[width=0.6\textwidth]{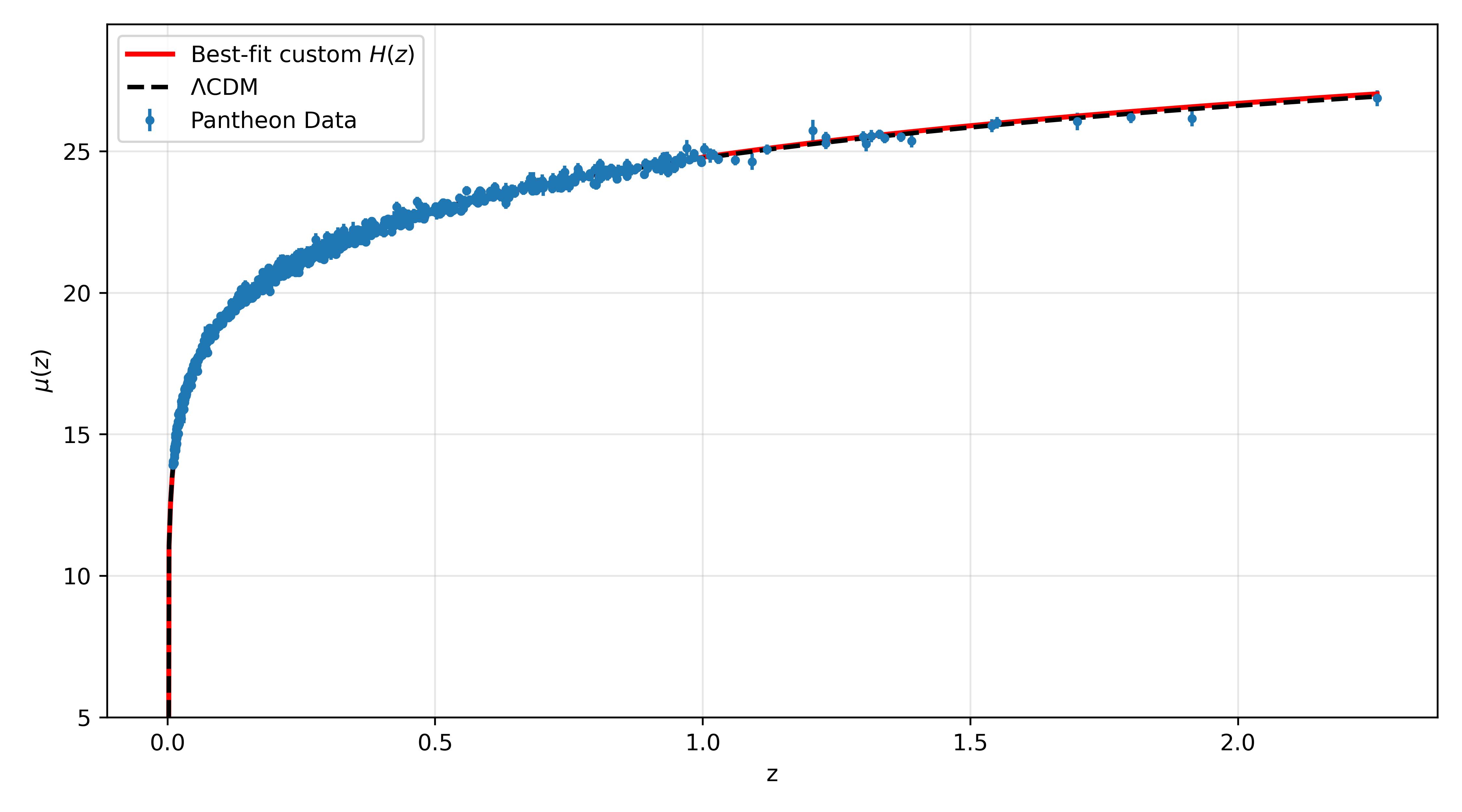} 
    \caption{Model(2) Pantheon data Error graph compare with $\Lambda\ CDM$ model}
     \label{figure:9}
\end{figure}

The Figure~\ref{figure:9} compares the observational data and Pantheon Type la Supernovae ( blue points with error bars ) with the theoretical distance modulus $\mu(z)$ read by the rebuilt cosmological model ( Solid red line ). For visual analogy, the average $\Lambda$ CDM model is described by the black dashed curve, which has been moved to match with data. Above the whole redshift range ($0 < z < 2.5$) , the rebuilt model and the observational data overlap nearly, indicating a good match and high degree of consistency with observational measurements, displaying that the indicated model successfully imitates the observed cosmic acceleration while keeping compliancy with the mainstream cosmological framework.

\begin{figure}[H]
    \centering
    \includegraphics[width=0.5\textwidth]{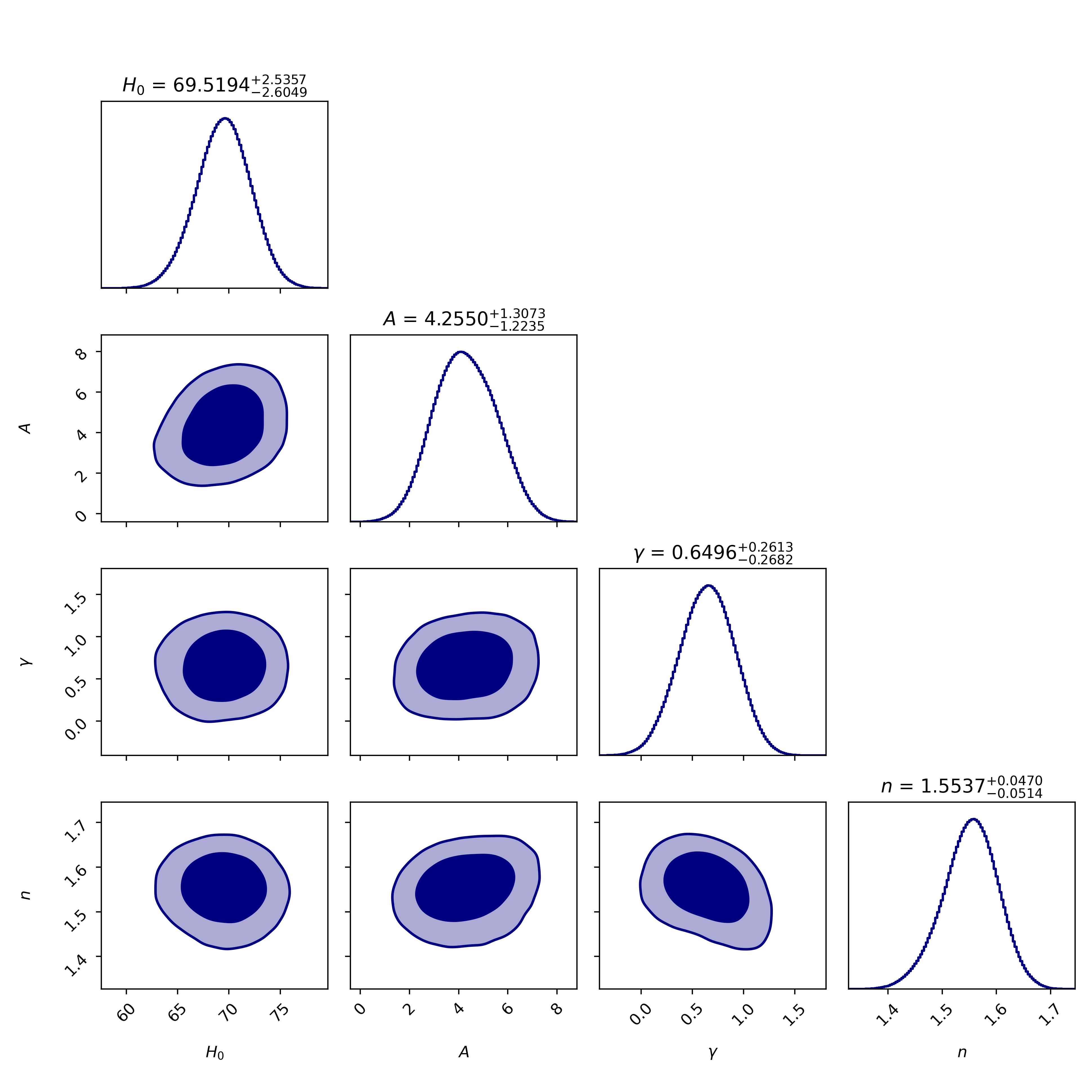} 
    \caption{Model(2) Combined parameter constraints from joint Hubble (OHD) and  Pantheon Type Ia Supernova datasets.}
     \label{figure:10}
\end{figure}

Applying a Markov Chain Monte Carlo (MCMC) investigation, the combined posterior distributions of the model parameters $H_0$, $A$, $\gamma$ and $n$ by using (Hubble+Pantheon) dataset are shown in the corner plot in Figure~\ref{figure:10}. The one dimensional marginalization probability allocations for each parameter are shown in diagonal panels simultaneously with their $68\%$ confidence interval. The plot  shows two confidence regions contours corresponding to $1\sigma$ ($68\%$ confidence level) and $2\sigma$ ($95\%$ confidence level), where inner navy blue area shows the $1\sigma$ uncertainity and outer light navy blue area shows $2\sigma$ uncertainity. The model selection statistics denotes $\mathrm{AIC}_{\text{model}} = 610.88$, $\mathrm{BIC}_{\text{model}} = 630.81$ and $\mathrm{DIC}_{\text{model}} = 605.17$. The best fit best-fit chi-square value is $\chi^{2} = 602.88$, resultant in a reduced chi-square $\chi^{2}_{\text{red}} = 0.562$. This shows that the model and observational data have great consensus.

\setlength{\extrarowheight}{2.5 pt}
\begin{table}[H]
\centering

\begin{tabular}{|l|l|l|}
\hline
\textbf{Data used} & \textbf{Parameters} & \textbf{Best fit values} \\ \hline

\multirow{4}{*}{Hubble} 
 & $H_0$ & $66.285^{+3.441}_{-3.450}$ \\ \cline{2-3}
 & $A$ & $4.425^{+2.369}_{-2.377}$ \\ \cline{2-3}
 & $\gamma$ & $0.024^{+0.571}_{-0.569}$ \\ \cline{2-3}
 & $n$ & $1.416^{+0.122}_{-0.122}$ \\ \hline

\multirow{4}{*}{Pantheon} 
 & $H_0$ & $67.1536^{+2.8312}_{-2.3646}$ \\ \cline{2-3}
 & $A$ & $1.8823^{+1.0420}_{-1.0657}$ \\ \cline{2-3}
 & $\gamma$ & $0.6264^{+0.3044}_{-0.3003}$ \\ \cline{2-3}
 & $n$ & $1.4787^{+0.0714}_{-0.0831}$ \\ \hline

 \multirow{4}{*}{Hubble+Pantheon} 
& $H_0$ & $69.5194^{+2.5357}_{-2.6049}$ \\ \cline{2-3}
 & $A$ & $4.2550^{+1.3073}_{-1.2235}$ \\ \cline{2-3}
 & $\gamma$ & $0.6496^{+0.2613}_{-0.2682}$ \\ \cline{2-3}
 & $n$ & $1.5537^{+0.0470}_{-0.0514}$ \\ \hline
\end{tabular}
\caption{Model(2) fits to cosmological data, showing dataset, parameter sets, and best-fit values.}
\label{tab:fit_stats3}
\end{table}

\begin{table}[H]
\centering
\resizebox{\textwidth}{!}{

\begin{tabular}{|c|c|c|c|c|c|c|c|c|}
\hline
\textbf{Data used} & $\chi^2$ & \textbf{Reduced} $\chi^2$ & \textbf{AIC} & \textbf{BIC} & \textbf{DIC} 
& $\boldsymbol{\Delta\mathrm{AIC}}$ 
& $\boldsymbol{\Delta\mathrm{BIC}}$ 
& $\boldsymbol{\Delta\mathrm{DIC}}$ 
\\
\hline
 Hubble&12.033  &0.463&20.03& 25.64& 12.56& 5.77& 9.97& 0.30 \\ 
\hline
 Pantheon&859.880&0.824& 867.880&887.695&857.162& 7.312&22.173& -1.406  \\ 
\hline
 Hubble+Pantheon&602.88&0.562&610.88& 630.81&605.17&2.81&12.78&1.10\\ 
 \hline
\end{tabular}}
\caption{Statistical results for different datasets for Model (2).}
\label{tab:fit_stats4}
\end{table}

\subsubsection{Case(III) : Model 3}

\begin{figure}[H]
    \centering
    \includegraphics[width=0.5\textwidth]{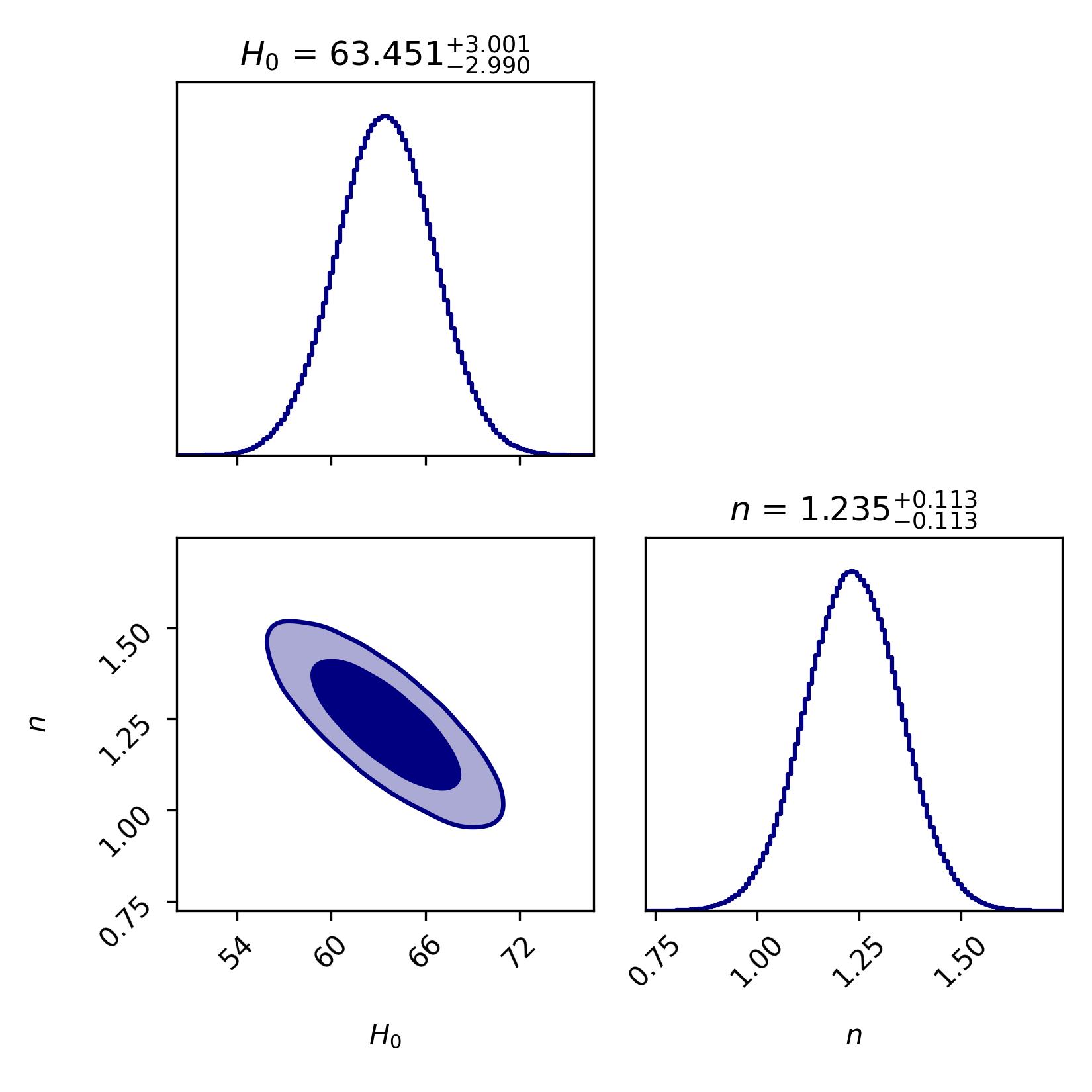} 
    \caption{Model(3) Confidence contours and marginalized posterior distributions for the parameters $H_0$ and $n$ obtained from the Hubble (OHD) dataset}
     \label{figure:11}
\end{figure}

The Hubble data was used to create contour maps for parameters ($H_0$, $n$) which is given in Figure~\ref{figure:11}.The lighter contours area show the 95\% levels, whereas the darker areas provide the 68\% confidence intervals. Tight constrains on the parameters are shown by the well defined and narrow peaks.  The model selection statistics denotes $\mathrm{AIC}_{\text{model}} = 15.05$, $\mathrm{BIC}_{\text{model}} = 17.85$ and $\mathrm{DIC}_{\text{model}} = 14.91$. The best fit best-fit chi-square value is $\chi^{2} = 11.050$, resultant in a reduced chi-square $\chi^{2}_{\text{red}} = 0.395$. This shows that the model and observational data have great consensus. In general, the findings indicates that the model gives a true and consonant chronology of the cosmological data.

\begin{figure}[H]
    \centering
    \includegraphics[width=0.6\textwidth]{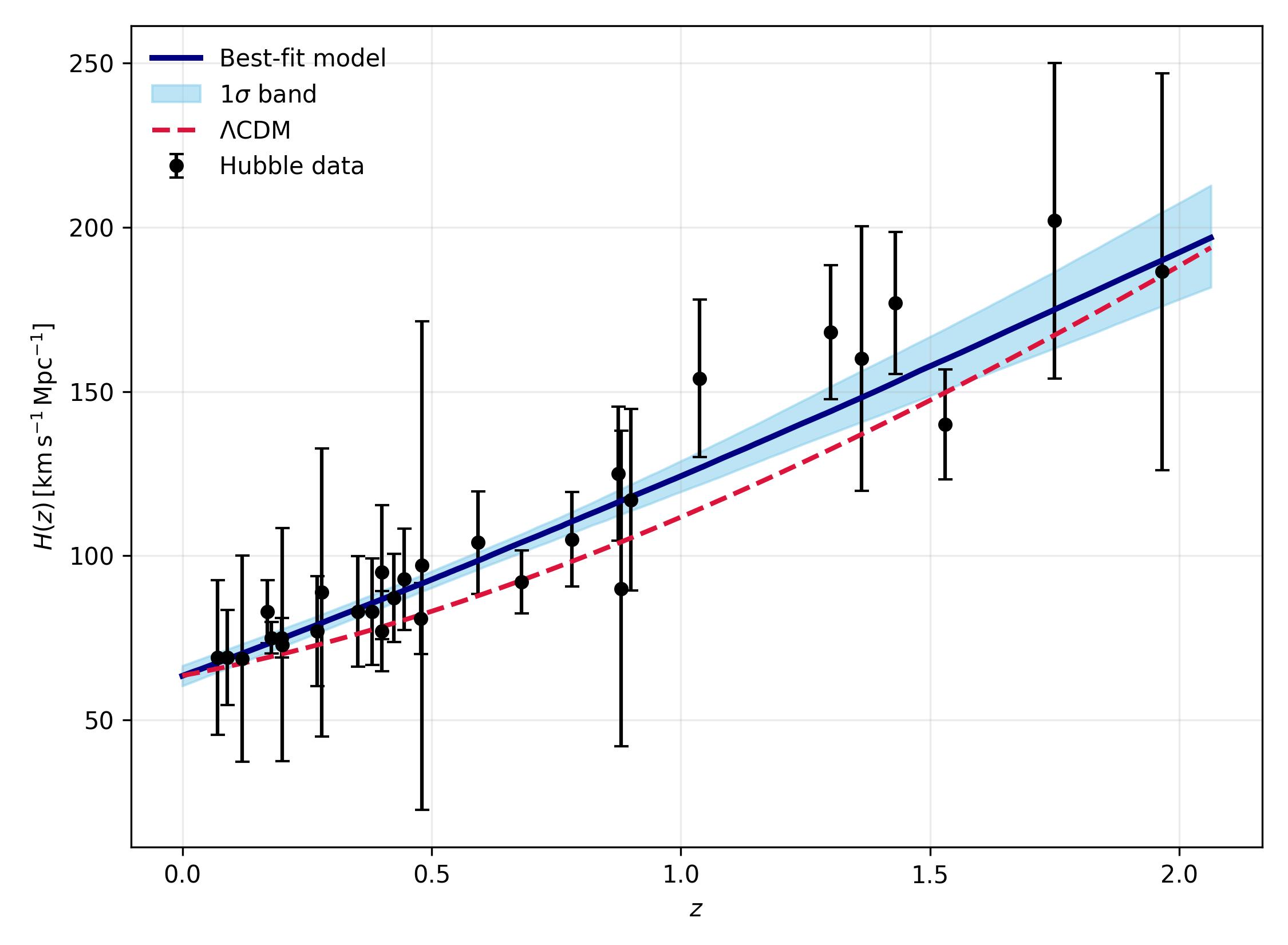} 
    \caption{ Model(3) Hubble data  Error graph compare with $\Lambda\ CDM$ model}
     \label{figure:12}
\end{figure}

Direct comparison with empirical Hubble parameter data can be used to evaluate the observational validation of model (3). Figure~\ref{figure:12} presents our Model's predictions , displayed next to the traditional $\Lambda$CDM model (red dashed curve). The graphic includes 30 observational Hubble data points from the redshift interval $0.07 < z < 2.0$, each of which is shown with an accompanying error bar. Our model's MAP best-fit accurately captures the overall pattern of the observational data, especially at low to intermediate redshift values. The model curve stays well within the observational uncertainties, despite minor variations that show up at higher redshifts, demonstrating the model's resilience to actual data. 

\begin{figure}[H]
    \centering
    \includegraphics[width=0.6\textwidth]{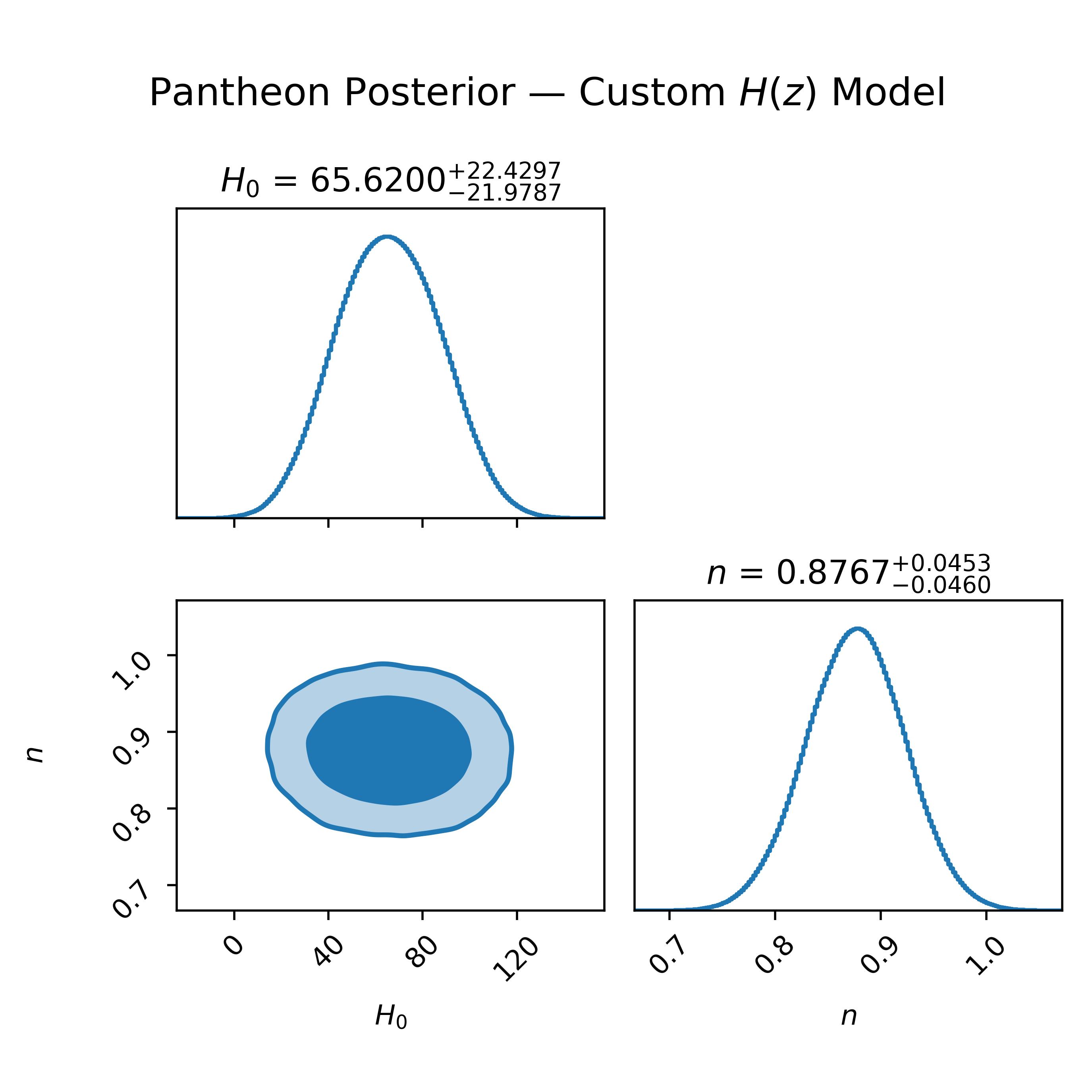} 
    \caption{Model(3) MCMC-based posterior distributions for parameters $H_0$ and $n$ using the Pantheon Type Ia Supernova dataset.}
     \label{figure:13}
\end{figure}

Applying a Markov Chain Monte Carlo (MCMC) investigation, the combined posterior distributions of the model parameters ($H_0$, $n$) by using Pantheon dataset are shown in the corner plot in Figure~\ref{figure:13}. The one dimensional marginalization probability allocations for each parameter are shown in diagonal panels simultaneously with their $68\%$ confidence interval. The plot  shows two confidence regions contours corresponding to $1\sigma$ ($68\%$ confidence level) and $2\sigma$ ($95\%$ confidence level), where inner blue area shows the $1\sigma$ uncertainity and outer light blue area shows $2\sigma$ uncertainity. The model selection statistics denotes $\mathrm{AIC}_{\text{model}} = 882.232$, $\mathrm{BIC}_{\text{model}} = 892.140$ and $\mathrm{DIC}_{\text{model}} = 880.234$. The best fit best-fit chi-square value is $\chi^{2} = 878.232$, resultant in a reduced chi-square $\chi^{2}_{\text{red}} = 0.840$. This shows that the model and observational data have great consensus.

\begin{figure}[H]
    \centering
    \includegraphics[width=0.6\textwidth]{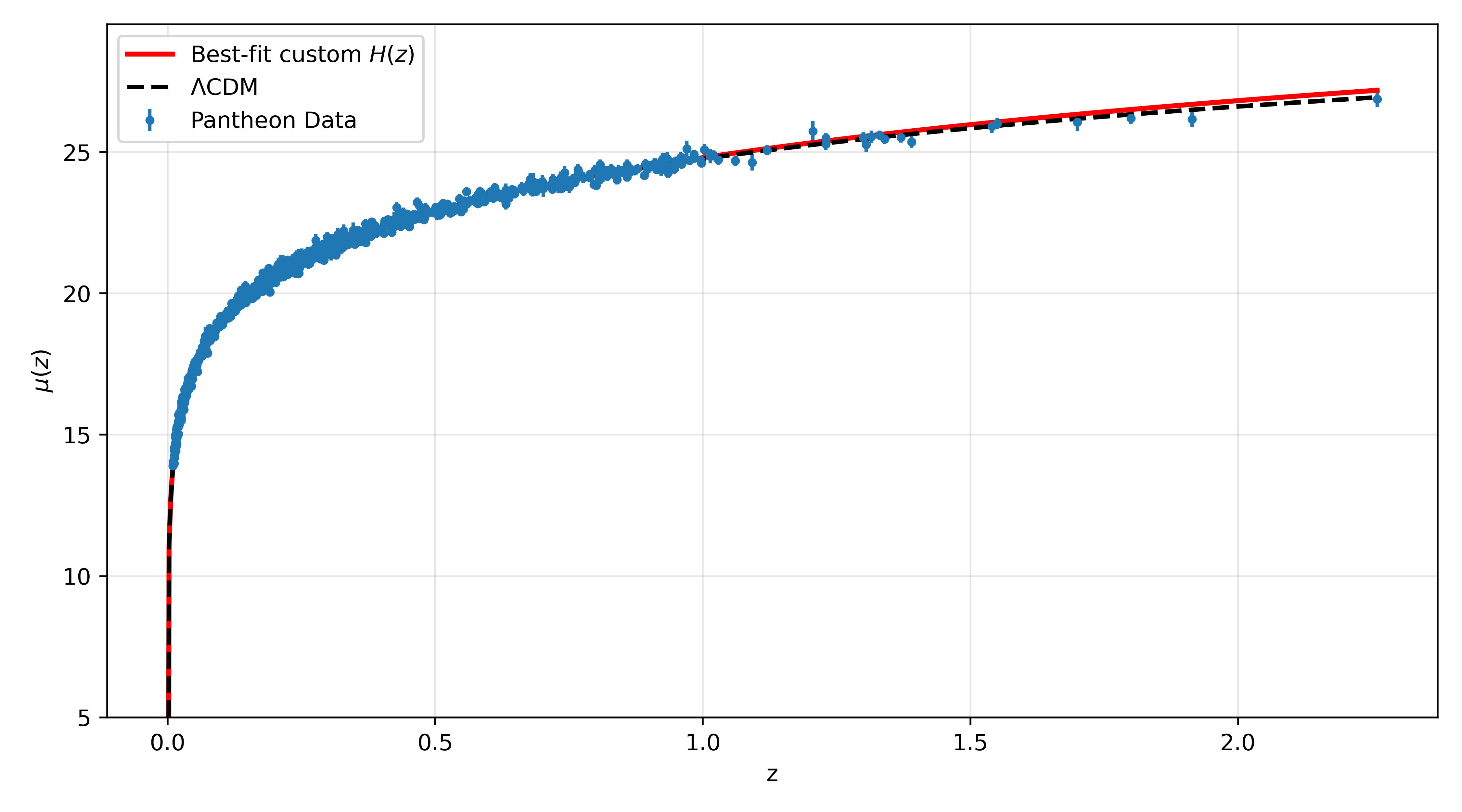} 
    \caption{Model(3) Pantheon data  Error graph compare with $\Lambda\ CDM$ model}
     \label{figure:14}
\end{figure}

The Figure~\ref{figure:14} compares the observational data and Pantheon Type la Supernovae ( blue points with error bars ) with the theoretical distance modulus $\mu(z)$ read by the rebuilt cosmological model ( Solid red line ). For visual analogy, the average $\Lambda$ CDM model is described by the black dashed curve, which has been moved to match with data. Above the whole redshift range ($0 < z < 2.5$) , the rebuilt model and the observational data overlap nearly, indicating a good match and high degree of consistency with observational measurements, displaying that the indicated model successfully imitates the observed cosmic acceleration while keeping compliancy with the mainstream cosmological framework.

\begin{figure}[H]
    \centering
    \includegraphics[width=0.6\textwidth]{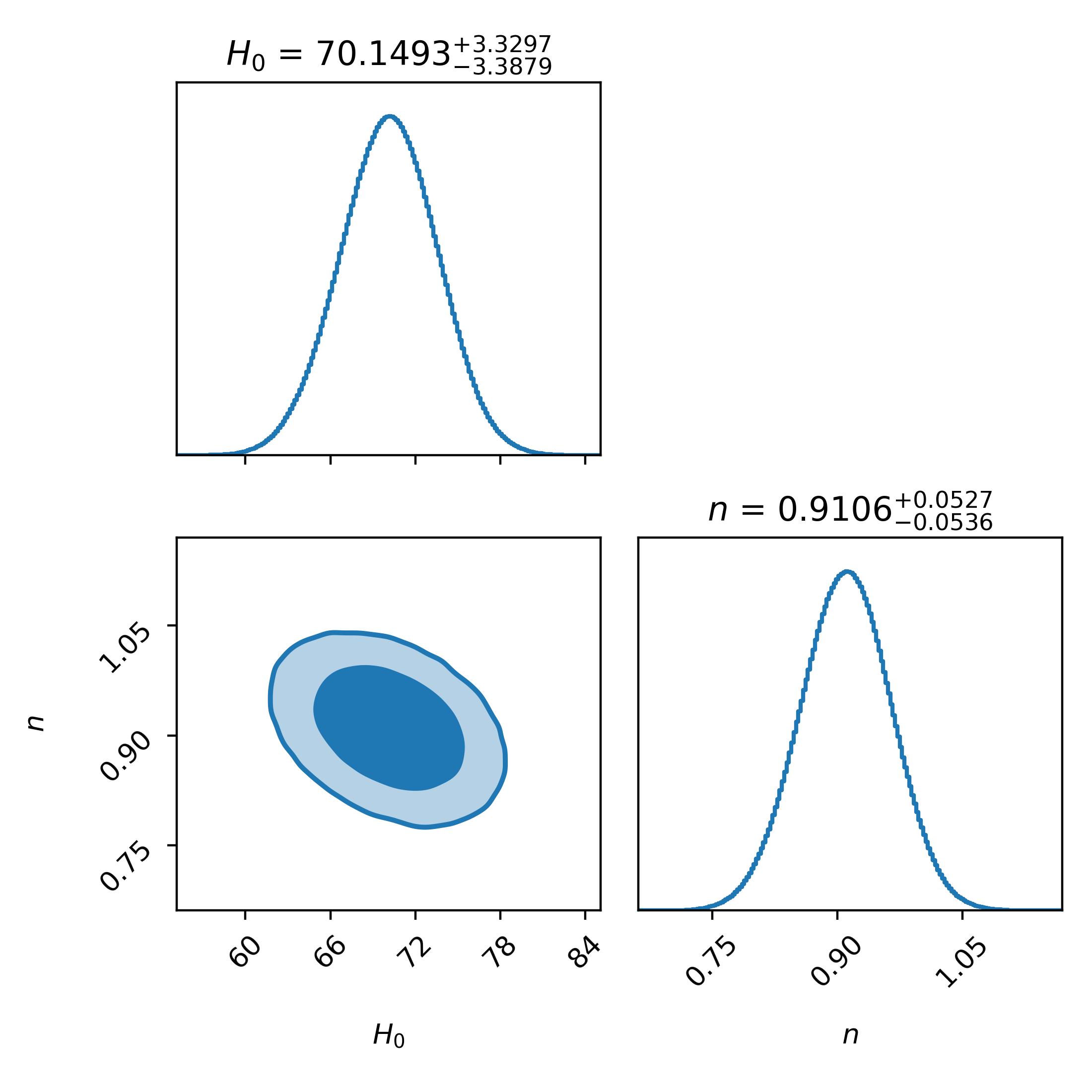} 
    \caption{Model(3) Combined parameter constraints from joint Hubble (OHD) and  Pantheon Type Ia Supernova datasets.}
     \label{figure:15}
\end{figure}

The (Hubble+Pantheon) data was used to create contour maps for  parameters ($H_0$, $n$) which is given in Figure~\ref{figure:15}. The lighter contours area show the 95\% levels, whereas the darker areas provide the 68\% confidence intervals. Tight constrains on the parameters are shown by the well defined and narrow peaks.  The model selection statistics denotes $\mathrm{AIC}_{\text{model}} = 632.87$, $\mathrm{BIC}_{\text{model}} = 642.84$ and $\mathrm{DIC}_{\text{model}} = 632.31$. The best fit best-fit chi-square value is $\chi^{2} = 628.87$, resultant in a reduced chi-square $\chi^{2}_{\text{red}} = 0.586$. This shows that the model and observational data have great consensus.

\setlength{\extrarowheight}{2.5 pt}
\begin{table}[H]
\centering

\begin{tabular}{|l|l|l|}
\hline
\textbf{Data used} & \textbf{Parameters} & \textbf{Best fit values} \\ \hline

\multirow{2}{*}{Hubble} 
 & $H_0$ & $63.451^{+3.001}_{-2.990}$ \\ \cline{2-3}
 & $n$ & $1.235^{+0.113}_{-0.113}$ \\ 
 \cline{2-3}
 \hline

\multirow{2}{*}{Pantheon} 
 & $H_0$ & $65.6200^{+22.4297}_{-21.9787}$ \\ \cline{2-3}
 & $n$ & $0.8767^{+0.0453}_{-0.0460}$ \\ \cline{2-3}
 \hline

 \multirow{2}{*}{Hubble+Pantheon} 
 & $H_0$ & $70.1493^{+3.3297}_{-3.3879}$ \\ \cline{2-3}
 & $n$ & $0.9106^{+0.0527}_{-0.0536}$ \\ \cline{2-3}
 \hline
\end{tabular}
\caption{Model (3) fits to cosmological data, showing dataset, model name, parameter sets, and best-fit values.}
\label{tab:fit_stats5}
\end{table}

\begin{table}[H]
\centering
\resizebox{\textwidth}{!}{

\begin{tabular}{|c|c|c|c|c|c|c|c|c|}
\hline
\textbf{Data used} & $\chi^2$ & \textbf{Reduced} $\chi^2$ & \textbf{AIC} & \textbf{BIC} & \textbf{DIC} 
& $\boldsymbol{\Delta\mathrm{AIC}}$ 
& $\boldsymbol{\Delta\mathrm{BIC}}$ 
& $\boldsymbol{\Delta\mathrm{DIC}}$ 
\\
\hline
 Hubble&11.050  &0.395 &15.05& 17.85& 14.91& -5.69& -4.29& -3.83 \\ 
\hline
 Pantheon&878.232&0.840& 882.232&892.140&880.234& 21.665&26.618& 21.666  \\ 
\hline
 Hubble+Pantheon&628.87&0.586&632.87& 642.84&632.31&-238.17&-233.19&-236.74\\ 
 \hline
\end{tabular}}
\caption{Statistical results for different datasets for Model (3).}
\label{tab:fit_stats6}
\end{table}

\begin{figure}[H]
    \centering
    \includegraphics[width=0.6\textwidth]{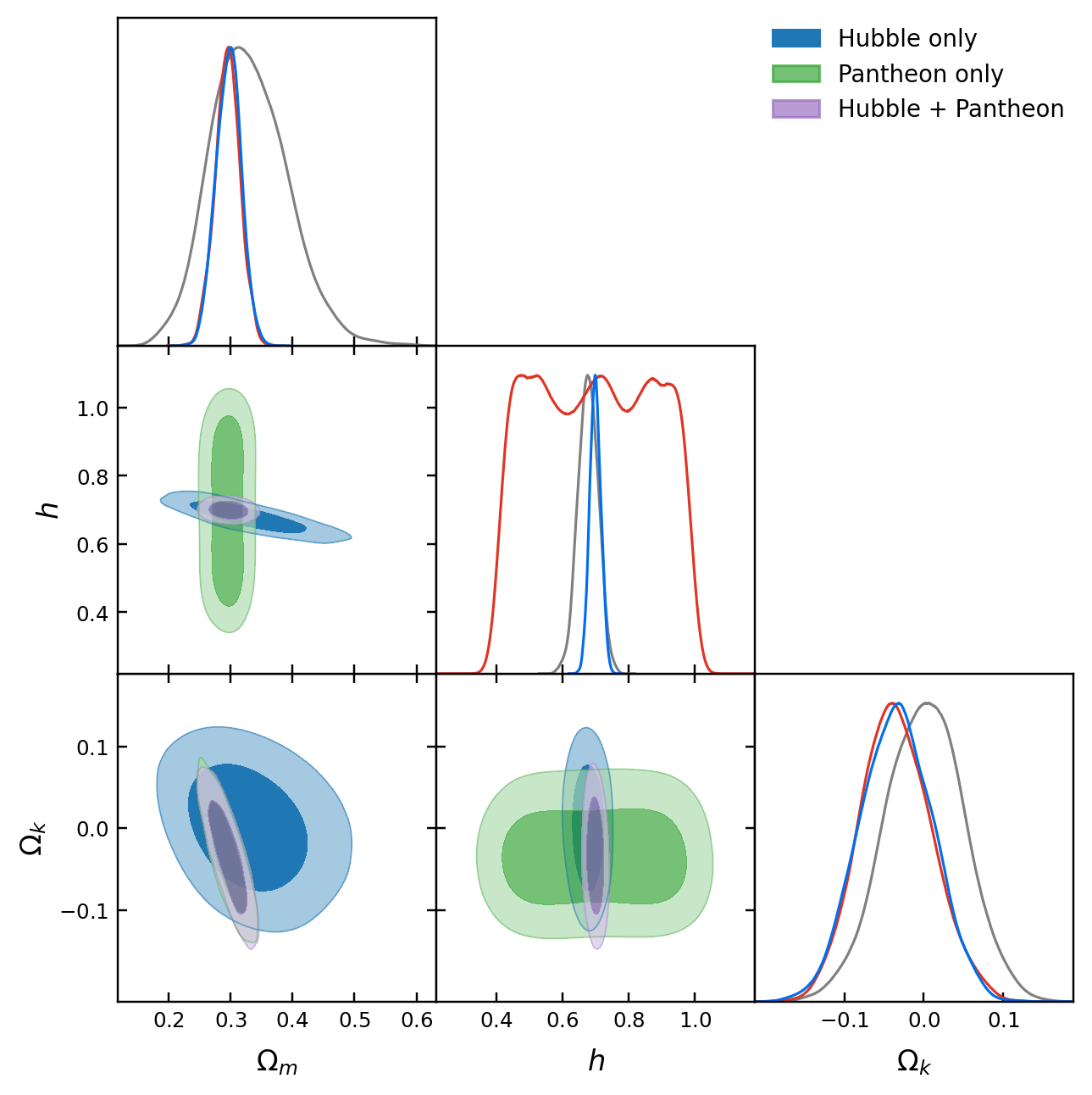} 
    \caption{$\Lambda CDM$ plot with hybrid dataset}
     \label{figure:16}
\end{figure}

The corner plot of cosmological parameter constraints derived from three datasets Hubble only (blue), Pantheon only (green), and the combined \textbf{Hubble+Pantheon} (purple) is shown in the  Figure~\ref{figure:16}.  A distinct and physically coherent tendency can be seen in the cosmological parameter values deduced from the three data combinations. The Hubble parameter is measured as $h = 0.6775 \pm 0.0307$ and the matter density is moderately constrained, $\Omega_m = 0.3307 \pm 0.0630$, using $\textbf{Hubble (cosmic chronometer) data only}$. Though, the curvature parameter is still weakly restrained, $\Omega_k = 0.0009 \pm 0.0500$, indicating that the sensitivity to spatial curvature. However, the constraint on the Hubble parameter is relatively weak, $h = 0.6980 \pm 0.1735$, because of the innate degeneracy between the supernova absolute magnitude and $H_0$. In contrast, the $\textbf{Pantheon supernova sample alone}$ offers a much tighter values on the matter density, $\Omega_m = 0.2953 \pm 0.0202$, indicating the strong geometric influence of luminosity distance measurements. Within uncertainties, the curvature parameter from Pantheon alone, $\Omega_k = -0.0346 \pm 0.0457$, is consistent with a nearly flat universe. The complementary nature of the two probes results in far better and more reliable restrictions when $\textbf{Hubble and Pantheon data are combined}$: $\Omega_m = 0.2970 \pm 0.0204$, $h = 0.6991 \pm 0.0171$, and $\Omega_k = -0.0347 \pm 0.0463$. In addition to confirming consistency with a spatially flat or slightly curved universe, the combined results show a clear tightening of the Hubble parameter uncertainty, emphasis the conclusiveness of joint investigations in breaking parameter degeneracies and enhancing cosmological parameter estimation.

\section{Hubble and deceleration parameters}

\begin{figure}[H]
    \centering
    \includegraphics[width=0.5\textwidth]{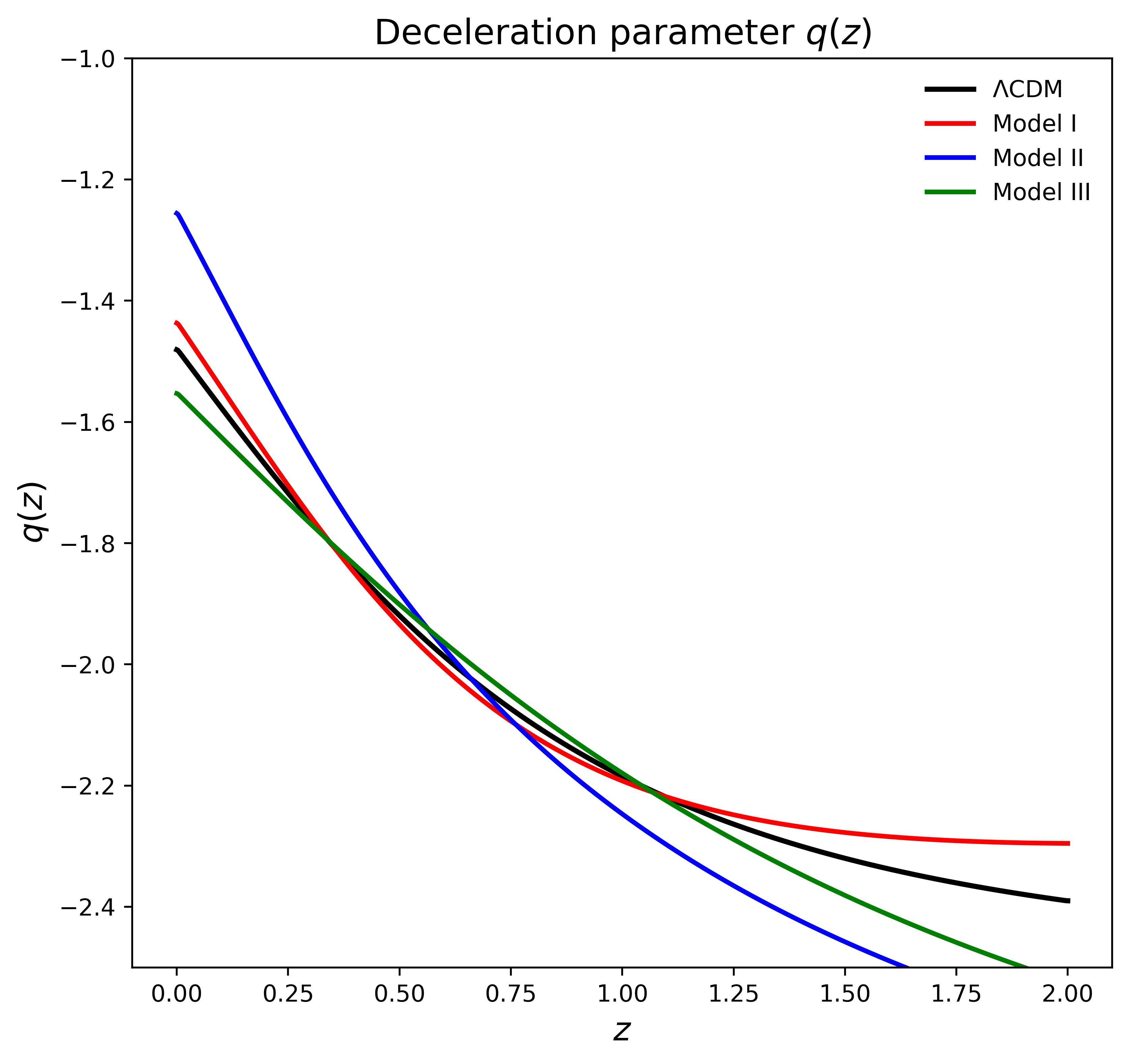} 
    \caption{Plot for Hubble and deceleration parameter $q(z)$}
     \label{figure:17}
\end{figure}
Using the linear–calibration fits $H_{\text{fit}}(z) = a\,H_{\text{model}}(z) + b$ reported in the output 
(Model I: $a = 115.02528977$, $b = -46.50737261$; 
Model II: $a = 35.63456955$, $b = 35.25176215$; 
Model III: $a = 25.8504532  $, $b = 41.73368892$; 
$\Lambda$CDM fit: $H_0 = 68.15718807$, $\Omega_m = 0.31948554$), 
we computed the deceleration parameter from the fitted Hubble function  \cite{chap02}
$$
q(z) = -\left(1 + \frac{\dot{H}}{H^2}\right)
      = -1 - \frac{1+z}{H(z)}\,\frac{dH}{dz},
$$
with $\frac{dH}{dz}$ evaluated numerically on a fine grid (central differences). 
Because for the linear calibration $H_{\text{fit}} = aH + b$ one has 
$\frac{dH_{\text{fit}}}{dz} = a\,\frac{dH}{dz}$, 
the fitted $q$-formula becomes
$$
q_{\text{fit}}(z) = -1 - (1+z)\,\left[\frac{a\,H'(z)}{a\,H(z) + b}\right],
$$
Consequently, the calibration shifts the denominator by $b$ while simultaneously rescaling the derivative. 
 In actuality, the specific $a,b$ numbers mentioned above do not result in a $q$ sign change. 
 We examined the displayed range $0 \le z \le 2$ . The fitted curves stayed negative everywhere we looked. 
 For comparison, over the depicted interval $0 \le z \le 2$. The fitted data show no acceleration-to-deceleration transition (no $q = 0$ crossing) since all of these values are absolutely negative as shown in Figure~\ref{figure:17}. Consequently, within the examined redshift range, all theories predict ongoing cosmic acceleration.

\section{Stability analysis}

The effective sound speed is used to analyze the models' dynamical stability and is defined as
$$
C_{s}^{2} \equiv \frac{d\bar{p}}{d\rho}, 
\qquad \bar{p} = p - 3\xi H ,
$$
with the stability 
$$
C_{s}^{2} \geq 0 \quad \text{(classical stability)}, 
$$
In terms of derivatives,
$$
C_{s}^{2} = \frac{dp}{d\rho} - 3\xi \frac{dH}{d\rho}.
$$

$C_{s}^{2}$ varies with cosmic time $t$ for three three models with bulk viscosity parameter $\xi = 0.65$

\begin{figure}[H]
    \centering
    \includegraphics[width=0.5\textwidth]{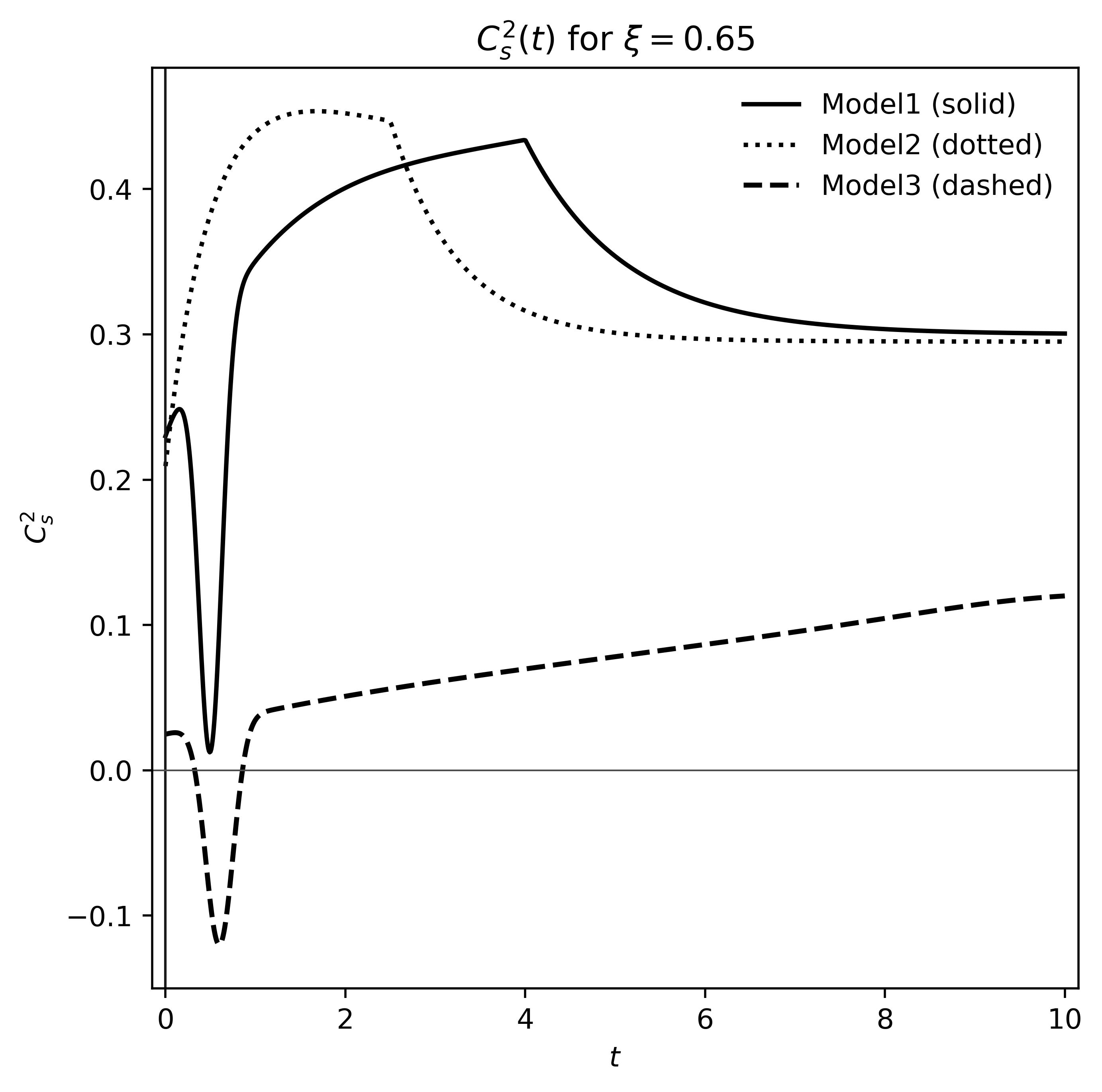} 
    \caption{Stability curve for Model (1),(2) and (3)}
     \label{figure:18}
\end{figure}
Since $C_s^2 > 0$ denotes a stable configuration and $C_s^2 < 0$ implies instability, Figure~\ref{figure:18} shows the evolution of the squared sound speed $C_s^2(t)$ for the three proposed cosmological models---Model~1 (solid), Model~2 (dotted), and Model~3 (dashed)---at a fixed viscous parameter $\xi = 0.65$. This serves as an indicator of their dynamical stability.  Model~1 first goes through an oscillating phase, as seen in the picture, with a transient instability ($C_s^2 < 0$) at early times ($t < 2$), before stabilizing around $C_s^2 \approx 0.3$ over time.  The strongest and most reliable stability of the three is shown by Model~2, which stays positive for the most of its history, peaking close to $t \approx 3$ and then moving toward a steady constant value. Model~3, on the other hand, begins with an unstable phase ($C_s^2 < 0$) around $t \approx 1$ and then progressively moves into a positive regime, eventually reaching $C_s^2 \approx 0.1$.  While Model~2 exhibits the highest level of stability, Model~1 achieves stability following early oscillations, and Model~3 stabilizes more slowly, all three models generally evolve toward stable configurations in the late universe. This indicates that the cosmological dynamics in each case asymptotically approach stable behavior at late cosmic times.

\section{Conclusion}

In this work, we have explored the dynamics of a bulk viscous cosmological model within the Friedmann–Robertson–Walker (FRW) spacetime framework, where the cosmic fluid obeys a nonlinear equation of state that unifies dark energy and dark matter behavior. The presence of bulk viscosity was incorporated into the Friedmann equations, modifying the background dynamics and influencing the Universe’s late-time acceleration. Analytical solutions for the energy density and Hubble parameter were obtained for different cases of the viscous coefficient, and these solutions were further examined through stability analysis and confrontation with observational data.

Our results show that the inclusion of a bulk viscous term plays a significant role in describing the accelerated expansion of the Universe without invoking an explicit cosmological constant. The derived expressions for the Hubble and deceleration parameters illustrate a smooth transition from an early decelerated phase to a late-time accelerated regime, consistent with modern observational evidence. Moreover, the effective sound speed analysis confirms that the proposed models remain dynamically stable and causal over the entire evolutionary range considered.

From the observational perspective, the models were tested using Hubble parameter (OHD) and Pantheon datasets, constrained through a Markov Chain Monte Carlo (MCMC) approach. The results demonstrate strong consistency with $\Lambda$CDM trends while allowing controlled deviations that may offer deeper insights into the physics of dissipative processes in the cosmic medium. The goodness-of-fit statistics, including the reduced $\chi^2$, AIC, BIC and DIC values, confirm that the viscous models provide a competitive description of the observed expansion history.

The statistical contours from the MCMC analysis also highlight well-defined parameter regions, revealing that the viscosity coefficient $\xi$ and the equation of state parameter $\gamma$ are tightly constrained by current data. This indicates that the bulk viscous framework is not only phenomenologically viable but also capable of capturing essential features of cosmic acceleration within a unified dark sector description.

For future research, several extensions of this work are promising. It would be valuable to generalize the analysis to non-flat universes ($k \neq 0$) and to consider a time-dependent or scale-dependent bulk viscosity coefficient. Additionally, studying the thermodynamic properties and entropy evolution in such viscous cosmologies could further illuminate the connection between cosmic expansion and gravitational thermodynamics. Another potential direction involves exploring the implications of these models in alternative gravity theories, such as $f(R)$, $f(Q,T)$, or mimetic gravity, where dissipative effects could arise naturally from modified geometric frameworks. Finally, confronting these extended viscous models \cite{ref37} with next-generation observational data, including CMB anisotropies and large-scale structure surveys, would provide a more robust test of their physical validity.

In summary, our findings reinforce the importance of bulk viscous cosmology as a compelling approach to explain the late-time acceleration of the Universe. By combining analytical modeling, dynamical stability analysis, and statistical inference with observational data, this study contributes to a deeper understanding of how dissipative processes can shape the evolution of the cosmic fluid and provide an alternative pathway to dark energy phenomenology.

\section*{Data Availability statement}
 This research did not yield any new data.

\section*{Conflict of Interest}
Authors declare there is no conflict of interest.

\section*{Funding}
This work was supported by the Deanship of Scientific Research, Vice Presidency for Graduate Studies and Scientific Research, King Faisal University, Saudi Arabia (KFU253534).

\section*{Acknowledgments}
MT gratefully acknowledges for JRF from  \textbf{DST-INSPIRE Fellowship(IF230556), Department of Science and Technology, Ministry of Science and Technology, government of India}. PKD would like to thank the Isaac Newton Institute for Mathematical Sciences, Cambridge, for support and hospitality during the programme Statistical mechanics, integrability and dispersive hydrodynamics where work on this paper was undertaken. This work was supported by EPSRC grant no EP/K032208/1. Also, PKD wishes to acknowledge that part of the numerical computation of this work was carried out on the computing cluster Pegasus of IUCAA, Pune, India and PKD gratefully acknowledges Inter-University Centre for Astronomy and Astrophysics (IUCAA), Pune, India for providing them a Visiting Associateship under which a part of this work was carried out.

\end{document}